\documentclass[twoside,english]{iopart}
\usepackage{geometry}
\usepackage{bbm}
\usepackage{amsbsy}
\usepackage{amstext}
\usepackage{graphicx}
\usepackage{subfig}
\usepackage{esint}
\usepackage[numbers]{natbib}

\makeatletter
\usepackage{iopams}
\usepackage{setstack}

\newcommand{\eqref}[1]{(\ref{#1})}
\usepackage{iopams}

\makeatother

\usepackage{babel}
\begin{document}
\title{Crossover scaling functions in the asymmetric avalanche process}	

\author{A.A. Trofimova$^{\ddagger,*,1}$ and A.M. Povolotsky$^{\dagger,\ddagger,2}$}
\address{{$^*$National Research University Higher School of Economics,
		20 Myasnitskaya, 101000, Moscow, Russia}}
\address{{$\dagger$Bogoliubov Laboratory of Theoretical Physics, Joint Institute
		for Nuclear Research, 141980, Dubna, Russia}}
\address{{$\ddagger$ Center for Advanced Studies, Skolkovo Institute of Science and Technology, Nobelya str.,
		1, 143026 Moscow, Russia}}
\eads{\mailto{$^1$nasta.trofimova@gmail.com}, \mailto{$^2$alexander.povolotsky@gmail.com}}

\begin{abstract}
We consider the particle current in the asymmetric avalanche process on a ring. It is known to exhibit a  transition from the intermittent to continuous flow at the critical density of particles. The exact expressions for the first two scaled cumulants of the particle current are obtained in the large time limit $t\to\infty$ via the Bethe ansatz and a perturbative solution of the TQ-equation. The results are presented in an integral form suitable for the asymptotic analysis   in the large system size limit $N\to\infty$. In this limit the first cumulant, the average current per site or the average velocity of the associated interface, is asymptotically finite below the critical density and grows linearly and exponentially times power law prefactor at the critical density and above, respectively.  
 The scaled second cumulant per site, i.e. the diffusion coefficient or the  scaled variance of the associated interface height,    shows the   $O(N^{-1/2})$ decay  expected for models in the Kardar-Parisi-Zhang universality class below the critical density, while it is growing as  $O(N^{3/2})$ and exponentially  times power law prefactor at the critical point and above. Also, we identify  the crossover regime  and obtain the scaling functions for the uniform asymptotics unifying the three regimes.  These functions are compared to the scaling functions describing  crossover of  the cumulants of  the avalanche size, obtained  as  statistics of the first return area under the time space trajectory of the Vasicek random process.   
\end{abstract}

\maketitle

\section{Introduction}	
The asymmetric avalanche process (AAP) is an interacting particle system related to  the asymmetric simple exclusion process (ASEP)  and the zero range process (ZRP) \cite{Liggett}. Its peculiar feature is the time-scale separation incorporated into the dynamics responsible for the appearance of instant non-local avalanche-like reconstructions of the system. 	AAP was introduced in \cite{PIPH2001} in an attempt of finding a Bethe ansatz solvable model with threshold avalanche dynamics that could develop a kind of the self-organized critical state \cite{BTW1987,Bak1996} characterized by giant avalanches spreading across the whole system. Considered in the conservative setting of the large periodic lattice AAP  was shown in \cite{PIPH2001} to exhibit a transition from the intermittent to continuous flow. It is marked by the divergence of the average stationary particle current in an infinite system as the density of particles approaches a critical value from below. The transition signalizes that the average avalanche size,  finite in the thermodynamic limit below the critical density,  grows with the system size above \cite{PPC2003}. 

The fluctuations of the particle current in the large time limit $t\to\infty$ are described in terms of its higher scaled cumulants, starting from the scaled variance or the diffusion coefficient. Their behaviour in the scaling limit characterizes the universality class the system belongs to \cite{Krug,Halpin-Healy_Zhang}. As an example, we mention the diffusion coefficient of a particle in the totally asymmetric simple exclusion process (TASEP)  on a ring. Derived first in \cite{DEM1993} with the help of the matrix product technique \cite{DEHP1993}, its power law dependence on the size of the system was one of the first direct confirmations of the Kardar-Parisi-Zhang (KPZ) universality \cite{KPZ} obtained from the exact solution.  Later the whole large deviation function containing infinitely many cumulants was also obtained for this case using the Bethe ansatz approach \cite{DL}. Its functional form obtained under the same scaling was conjectured to be universal within the KPZ universality class. This universality was later tested with plenty of other examples (ASEP \cite{LK}, q-Boson ZRP \cite{Povolotsky2003,PM2006}, AAP \cite{PPC2003_AAP}, directed polymer in random medium \cite{BD2000}, e.t.c.), and the results on current large deviations were also extended to the system with open boundary conditions (TASEP \cite{LM2011}, ASEP \cite{GE, LMV2012, LP2014}). 

Also, when transitions between different types of scaling of fluctuations are considered, the diffusion coefficient provides a universal scaling function connecting the two regimes. For instance, the diffusion coefficient of a particle in ASEP considered as a function of asymmetry of particle hopping yields the universal scaling function describing the crossover between the KPZ and Edward-Wilkinson (EW) \cite{EW} regimes.   
For the first time it was obtained in \cite{DM} via matrix product ansatz and then rederived in \cite{PM2008} using the Bethe ansatz and Baxter's TQ-equation method \cite{Baxter1}. Its universality was later tested with an example of the model of q-boson ZRP \cite{TP2020}.

For the AAP model it was shown in \cite{PPC2003_AAP} that below the critical density the fluctuations of particle current developing in the KPZ scale are characterized by the large deviation function from \cite{DL}. However,  at the critical point and above, when the current respectively grows linearly and exponentially with the system size, the fluctuations are expected to be different from the KPZ ones. Under the mapping of 1+1 dimensional particle system to a 2+1 dimensional interface this change in the behaviour was associated in \cite{PPC2003}  with the depinning transition suffered by an interface tilted away from the hard direction in the random medium with anisotropic quenched disorder \cite{TKD}.  Below the depinning threshold, the thermal fluctuations cause finite avalanches, responsible for infinitely slow creep of the interface \cite{LT}. This slow motion is characterized by the KPZ fluctuations in an appropriate slow time scale. The avalanches grow unboundedly at the depinning threshold producing the disturbance front rapidly moving through the system with constant velocity \cite{K98}. The behaviour of this front was also predicted to be described by the KPZ universality class in the reduced by one spacial dimensionality, which corresponds to the simple 1+1-dimensional diffusion in our case.

While the scaling function describing the crossover of the average particle current at this transition was already discussed in \cite{PPC2003}, neither the functions for the higher cumulants having to do with the crossover of fluctuations nor even the scale of this crossover are  yet known. Note that the technique used in \cite{PPC2003_AAP} for obtaining the higher current cumulants was based on the asymptotic method of solution of the Bethe ansatz equations developed in \cite{KIM, LK} that suggests a  special KPZ-specific  scaling. Therefore, it is not suitable for the analysis of current fluctuations in and above the critical point, which thus have remained beyond the scope of analytic tools to date. Here, we apply an alternative technique based on  Baxter's TQ-equation \cite{Baxter1}, developed in \cite{PM2008} for studying current cumulants in the ASEP. This method allows obtaining exact formulas for the average current and the diffusion coefficient in the form of sums of contour integrals, similar to those obtained previously in \cite{TP2020} for the q-boson ZRP.   Then, we perform the asymptotic analysis based on the saddle point approximation, similar to the one that was used in \cite{TP2020}, though modified to take into account the effect of a  pole of the integrand,  approaching the saddle point and crossing the steepest descent contour, when the transition takes place. 
The asymptotic analysis performed in the thermodynamic limit yields different power law dependencies of the diffusion coefficient on the system size below and at the critical point and an exponential growth above. We also perform the analysis in a special scaling limit, when the deviation of the particle density from its critical value vanishes as the system size grows. This yields uniform asymptotics unifying the three regimes, producing a new crossover scaling function.

Then, we compare the scaling functions describing the crossover of the time integrated particle current    with the  scaling functions describing the crossover of the avalanche size. Specifically, the particles in  AAP are brought by avalanches, i.e. the number of particle jumps for some period is nothing but the sum of sizes of avalanches that happened for this period. One could expect that the behaviour of the avalanche size is similar to that of the time integrated particle current normalized by the number of avalanches.  However, the avalanches at different time moments are correlated. Therefore, a direct correspondence between the particle current and the avalanche size holds only on the level of the first moment, for which the correlations are irrelevant.   The discrepancy of the second cumulant characterizes the effect of correlations on the squared sum of avalanche sizes. Here, we study the avalanche size in the scaling regime, in which the uniform asymptotics of the current cumulants unifying the three regimes was obtained. It is natural to expect that the large scale physics in the close vicinity of the phase transition have a continuous description, i.e. in our  case can be reduced to some stochastic PDE.  Indeed, the number of particles involved in an avalanche at every step of the discrete time dynamics performs a kind of a biased random walk with transition probabilities defined in terms of the stationary state of the discrete version of the AAP,  reaching zero when the avalanche ends. We argue that in the scaling limit the rescaled process converges to a version of the Ornstein-Uhlenbeck process \cite{OU, Doob} with an additional constant drift term, known as Vasicek model \cite{V}. Then, the rescaled avalanche size is the area under the space-time trajectory of  this process before its first return to the origin. Generalizing  recent result  \cite{KM2021}  on   the  statistics of the first passage area functional of the  Ornstein–Uhlenbeck process  we find the first two cumulants of the first return area in the Vasicek model as a function of the additional drift  related to the rescaled deviation of the AAP particle density  from its critical value. As a result, we obtain the same crossover function for the first moment of the avalanche size as for the particle current. For the second cumulant, the expressions are different. They approach the same leading asymptotics far above the critical point, while at low densities the correlations are dominant, responsible for the KPZ scaling of the diffusion coefficient. 	

Our article is organized as follows.
In section \ref{sec: AAP:Model and results} we introduce the AAP model and the current cumulants. We list the main results of the article consisting of exact integral expressions for two first current cumulants, their asymptotic behaviour in the thermodynamic limit and their uniform asymptotic formulas unifying the three regimes. The last ones are formulated in terms of the crossover scaling functions. In section \ref{sec: Crossover and first passage} we find the crossover functions for the first two cumulants of the avalanche size and compare them to those for the average current and diffusion coefficient. In the remaining part of the article, we outline the calculations. We obtain exact formulas using the Bethe ansatz and TQ-equation methods in section \ref{sec: TQ} and we perform the asymptotic analysis in section \ref{sec: Asymptotic analysis}. In the appendix, we sketch  
the derivation of the average current using the averaging over the stationary state of the discrete time version of the AAP following mainline of \cite{PPC2003}.

\ack{}{}
The work of A.Trofimova was prepared within the framework of the Basic Research Program at HSE University. The work of A.Povolotsky in part of the results of sections 3-5 has been funded by the Russian Science Foundation under grant 19-11-00275 via  Skolkovo  Institute  of  Science  and  Technology.

\section{Asymmetric Avalanche Process: Model and results} \label{sec: AAP:Model and results}

\subsection{Model and its dynamics}

The asymmetric avalanche process(AAP) is a one-dimensional stochastic interacting particle system
formulated as a continuous time Markov process   $\mathbf{n}(t):\mathbb{R}_{\geq0} \to \Omega$ on a state space  $\Omega = \{0,1\}^{\mathbb{Z}/N\mathbb{Z} }$ consisting of particle configurations on a periodic one-dimensional lattice   with $N\in \mathbb{N}$ sites  (sites $i$ and $N+i$ are identical), where every site can accommodate no more than one particle. Such configurations will be referred to as \textit{stable}. 

Having started with a stable configuration the subsequent evolution proceeds as follows.

\begin{figure}[h]%
	\centering
	\subfloat[ ]{{
			\includegraphics[width=7cm]{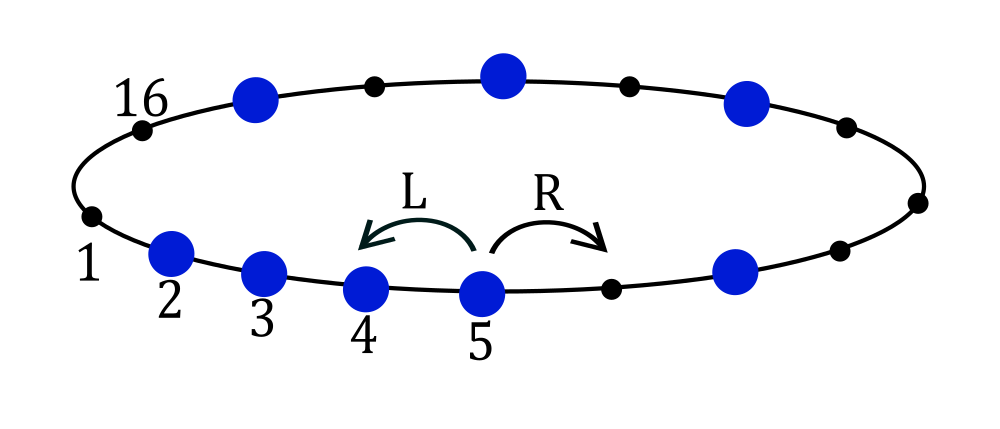} }}%
	\qquad
	\subfloat[ ]{{\includegraphics[width=7cm]{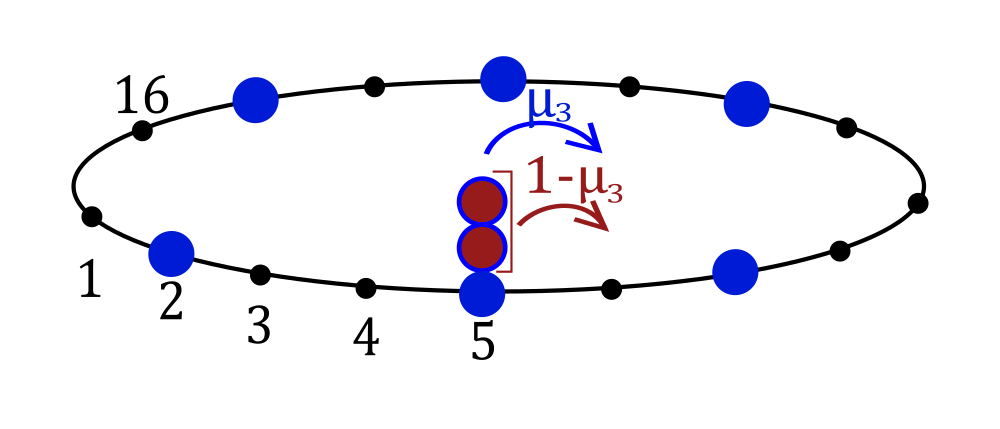} }}%
	\caption{Poissonian (a) and discrete time avalanche (b) dynamics of AAP  on a ring with $N=16$ sites and $p=8$ particles. (a) Poissonian jump of the particle at position $5$ out of a stable configuration can be made left with rate $L$ or right with rate $R$. (b) Step of the avalanche dynamics,  in which either three particles from the active site $5$ jump one step right all together with probability $\mu_3$ (blue) or  two  particles jump to the right with the probability $1-\mu_3$ (red). }%
	\label{fig:AAP}%
\end{figure}

\textit{Poissonian jumps.}  Every particle is assigned an  independent exponential clock with rate one.  When a particle's clock rings, the particle jumps to the neighboring left or right site choosing them with probabilities $L$ and $R$ respectively,  Fig.~\ref{fig:AAP}(a). If the neighboring site is empty, the jump results in a  particle exchange between the two neighboring sites leading to another stable configuration. The jump towards an occupied site leads to another stable configuration that may be a  highly non-local rearrangement of the configuration of departure. The route between the two configurations can be recast in terms of an instant discrete time avalanche.

\textit{Avalanche dynamics.} An avalanche develops in discrete time   passing through a series of  \textit{unstable} particle configurations, in which all sites  contain at most  one  particle except for one site, referred to as \textit{active}, that contains  $n \geq 2$ particles. It starts, when  the active site with $n=2$ has appeared after a particle made a Poissonian jump to a neighboring  occupied site. If at any step of the avalanche the site $i$ is active with  $n\geq 2$ particles, then 
\begin{itemize}
	\item either all $n$ particles move from site $i$ to site $i+1$ with probability $\mu_{n}$ 
	\item  or $n-1$ particles move from site $i$ to site $i+1$ with probability $1-\mu_n$ and one particle remains at $i$, Fig.\ref{fig:AAP}(b),
\end{itemize}
where $\{\mu_n\}_{n\in\mathbb{N}}$ is in general a fixed set of arbitrary probabilities, $0\leq \mu_n< 1.$ 
The avalanche ends, when one particle jumps from the active site with $n=2$ particles to an empty site and the system arrives at a stable  configuration. From the point of view of the  Poissonian clocks  the avalanche is instant, i.e. is not visible in the continuous time scale   and enters only via the  rates of transitions between stable configurations.

A particle configuration is specified by the list  of occupation numbers $\mathbf{n}=(n_1,\dots,n_N),$ where $n_i\in \{0,1\}$ for $i=1,\dots,N$. To describe the evolution of the system, we use the probability 
\begin{equation}
P_t(\mathbf{n})=\mathbb{P}(\mathbf{n}(t)=\mathbf{n})
\end{equation}   
for  the system to be in configuration $\mathbf{n}$ at time $t$. Given  the initial distribution $P_0(\mathbf{n})$, it  solves the master equation 
\begin{equation}
\partial_{t}P_{t}(\mathbf{n})=\mathcal{L}P_{t}(\mathbf{n}),\label{ME}
\end{equation}
where the generator $\mathcal{L}$ is a linear operator defined as
\begin{equation}
\mathcal{L}P_t(\mathbf{n})=\sum_{\mathbf{n}\in \Omega:\mathbf{n'}\neq\mathbf{n}}\left(u_{\mathbf{n}',\mathbf{n}}P_t(\mathbf{n'})-u_{\mathbf{n},\mathbf{n}'}P_t(\mathbf{n})\right),\label{eq: generator}
\end{equation}
where $u_{\mathbf{n}',\mathbf{n}}$ is the rate of transition  from configuration $\mathbf{n}'$ to configuration $\mathbf{n}$ following a Poissonian jump.

There is a special  $\textit{integrable}$ choice of toppling probabilities  \cite{PIPH2001}, given by a one-parametric family  
\begin{equation}
\mu_n= 1 - [n]_{q}, \quad \text{where} \quad [n]_q := \frac{1-q^n}{1-q}, \quad -1<q<0
\label{integr_cond}
\end{equation}
that allows one to write the master equation in relatively simple manageable form.  

To this end, we first note that the dynamics described preserves the number of particles in the system $||\mathbf{n}||=n_1+\cdots+n_N$. Let us fix this number to be 
\begin{equation}
||\mathbf{n}||=p.\label{eq: n=p}
\end{equation} 
Then, we use an alternative representation of stable particle configurations. Instead of specifying the occupation  numbers $n_1,\dots,n_N$ of all sites, we will list the coordinates of  particles   in the increasing order 
\begin{equation}
\boldsymbol{x}=\{x_{1}<,\dots <,x_{p} \}. \label{eq: domain}
\end{equation}
The periodic boundary conditions imply that this set is extended to the countable ordered set $\{x_i\}_{i\in{\mathbb{Z}}}$ under assumption   $x_{i+p}=x_i+N$. Then, the action  $\mathcal{L}P(\boldsymbol{x})$ of the generator (\ref{eq: generator}) is as follows. 
When $\boldsymbol{x}$ is a stable configuration without two particles occupying two neighboring sites, i.e.  $x_{i+1} - x_i > 1,$ $ i = 1, \dots, p$, $\mathcal{L}$ acts as a generator of independent Poissonian random walks,  
\begin{equation}
\mathcal{L}P_{t}(\boldsymbol{x})= \sum_{i=1}^p \left(L P_{t}
( \boldsymbol{x}_i^+) + R P_{t}
(\boldsymbol{x_i^-})\right)- p P_{t}(\boldsymbol{x}), \label{eq: master free}
\end{equation}
where for the brevity we denote $\boldsymbol{x}_i^+$($\boldsymbol{x_i^-}$) the state with $x_i$ increased(decreased) by $1.$ The other cases describe the interactions involving avalanches.   
With toppling probabilities (\ref{integr_cond})  they all can be dealt with in one go by extending   eq. (\ref{eq: master free})  to the whole domain (\ref{eq: domain}) with boundary conditions 
\begin{equation}
P_t(\dots, x, x,\dots) = (1+q) P_t(\dots, x-1, x,\dots)-q P_t(\dots,, x-1, x-1,\dots).
\label{eq: boundary_cond}    
\end{equation}
imposed to re-express  the terms beyond the domain (\ref{eq: domain}) via those inside. Note that since the unallowed terms appear in both sides of eq.~(\ref{eq: boundary_cond}) one should apply these boundary conditions recursively to obtain an infinite sum of terms corresponding to avalanches via which one can   reach given stable state constructed with the one-site toppling probabilities (\ref{integr_cond}). The fact that all the many-particle interactions emerge from the two-particle boundary conditions ensures applicability of the  coordinate Bethe ansatz to diagonalization of the generator of   AAP model.  
	

\subsection{Model observables\label{sec: Model observables}}
The observable of our interest is a total distance $Y_{t}$ traveled by all particles including  particle jumps during avalanches for the time $t$ passed since the beginning of the AAP evolution at $t=0$.  To describe the statistics of  $Y_{t}$ at large time we, as usual,  introduce the joint probability $P_{t}(\mathbf{n};Y)=\mathbb{P}(\mathbf{n}(t)=\mathbf{n},Y_t=Y)$ for the system to be in configuration $\mathbf{n}$  having the distance traveled by particles  $Y_{t}=Y$ at time $t$. Given initial conditions $P_{0}(\mathbf{n};Y)=P_{0}(\mathbf{n})\delta_{Y,0},$  its  generating function 
$$
G_{t}(\mathbf{n};\gamma)=\sum_{Y=-\infty}^{\infty} P_t(\mathbf{n};Y)e^{\gamma Y}
$$
solves the deformed version of the master equation \eqref{ME},
$$
\partial_{t}G_{t}(\boldsymbol{x};\gamma)=\mathcal{L_{\gamma}}G_{t}(\boldsymbol{x};\gamma),
$$
where  the operator $\mathcal{L}_{\gamma}$, 	
\begin{equation}
\mathcal{L}_\gamma G_t(\mathbf{n};\gamma)=\sum_{\mathbf{n}\in \Omega:\mathbf{n'}\neq\mathbf{n}}\left(u_{\mathbf{n}',\mathbf{n}}(\gamma)G_t(\mathbf{n'};\gamma)-u_{\mathbf{n},\mathbf{n}'}G_t(\mathbf{n;\gamma})\right),\label{eq: generator deformed}
\end{equation} 
is a non-stochastic deformation of  $\mathcal{L}$ obtained from the latter by   change  of   the off-diagonal matrix elements 
to those with the deformed rates 
\begin{equation}\label{eq: deformed rates}
u_{\mathbf{n}',\mathbf{n}}(\gamma)=\left.\frac{d}{dt}\right|_{t=0}\mathbb{E}\left(\mathbbm{1}_{\mathbf{n}(t)=\mathbf{n}}e^{\gamma Y_t}\right|\mathbf{n}(0)=\mathbf{n}').
\end{equation}
having a meaning of generating function of moments of the number of particle jumps within an instant transition from configuration $\mathbf{n}'$ to $\mathbf{n}$. Similarly to (\ref{eq: master free},\ref{eq: boundary_cond}),  the action (\ref{eq: generator deformed})  corresponding to the toppling probabilities (\ref{integr_cond}) can be written in $\boldsymbol{x}$-representation of  particle configurations as a deformation of the  free part 
\begin{eqnarray} \nonumber
\mathcal{L}_{\gamma}G_{t}(\boldsymbol{x}; \gamma)= \sum_{i=1}^p \left(Re^{\gamma} G_{t}(
\boldsymbol{x}_i^-;\gamma)+L e^{-\gamma} G_{t}
(\boldsymbol{x}_i^+; \gamma)\right)- p G_{t}(\boldsymbol{x};\gamma), \nonumber
\end{eqnarray}
with the deformed boundary conditions 
\begin{equation}
\hspace{-0.1\textwidth}	G_t(\dots, x,x, \dots ;\gamma ) = (1+q) e^{\gamma}G_t(\dots, x-1,x,\dots;\gamma) - q e^{2\gamma} G_t(\dots, x-1,x-1,\dots; \gamma).
\end{equation}
The generating function $G_{t}(\boldsymbol{x};\gamma)$
can be used to write  the moment generating function of the random variable $Y_{t}$  
\begin{equation}
\mathbb{E}e^{\gamma Y_{t}}=\sum_{\mathbf{n}\in\Omega}G_{t}(\mathbf{n};\gamma).
\end{equation}
The behaviour of the moment generating function in the large  time limit, $t\rightarrow \infty,$ is dominated by the largest eigenvalue $\lambda(\gamma)$ of the matrix $\mathcal{L_{\gamma}},$
\begin{equation}
\lambda(\gamma)=\lim_{t\rightarrow\infty}\frac{\ln\mathbb{E}e^{\gamma Y_{t}}}{t}=\sum_{n=1}^{\infty}c_{n}\frac{\gamma^{n}}{n!},
\end{equation}
where the second equality emphasizes that the function $\lambda(\gamma)$ is the   generating function of scaled cumulants 
\begin{equation}
c_{n}=\lim_{t\to\infty}\frac{\big\langle Y_{t}^{n}\big\rangle_{c}}{t},
\end{equation}
of  the total current $Y_t$,  where we use notation $\big\langle\xi^{n}\big\rangle_{c}$
for  order $n$ cumulant of the random variable $\xi$. In the present paper we deal with the first two scaled cumulants,  mean integral particle current 
\begin{eqnarray*}
	J=J(N,p) &  =&c_{1}=\lambda'(0) \label{eq: lambda'(0)}
\end{eqnarray*}
and group diffusion coefficient
\begin{eqnarray*}
	\Delta=\Delta(N,p) &  = &c_{2}=\lambda''(0).\label{eq: lambda''(0)}
\end{eqnarray*}
Below we will be interested in the behaviour of these quantities in the \textit{thermodynamic} limit of large system size with the density 
$$\rho=p/N$$ fixed. Then, it is  more convenient to use  the intensive versions of  current and  diffusion coefficient, the current 	and  diffusion coefficient per  site  
\begin{equation}
j_{N}(\rho)=\frac{J}{N},\quad	\Delta_{N}^j(\rho)=\frac{\Delta(N,p) }{N^2},
\end{equation}
which in particular can be  translated to the local statistics of the interface height under the particle system-interface mapping.

Before going to the results, let us discuss the relationship between the current cumulants and the statistics of avalanches. 	The random variable $Y_t$ is a functional on the trajectories of the process that can be represented in the following form. Let $\mathfrak{N}_t(p)$ be the Poisson process with the arrival rate $p$ that counts the number of Poissonian jumps of particles in the system for the time $t$. Then 
\begin{equation}\label{key}
Y_t=\sum_{i=1}^{\mathfrak{N}_t(p)}S_i,
\end{equation} 
where the random variables $S_i, i\in \mathbb{N}$, are the signed avalanche sizes, i.e. distance traveled by all particles within an avalanche following (and including) the Poissonian jump number $i$ at time $t_i$. 
Note that the sequence of configurations $\{\mathbf{n}(t_i)\}_{i\in \mathbb{N}}$  visited by the process being itself a discrete time Markov chain with the same stationary distribution as the continuous time one is independent of waiting times between the jumps and, hence, of the whole process $\mathfrak{N}_t(p)$.
Also, each random variable $S_i$ depends on the history of the continuous time process only via the particle configurations the avalanche number $i$ starts and ends at. In particular, it is independent of $\mathfrak{N}_t(p)$ and the conditional expectation $\mathbb{E} \left(S_i|\mathbf{n}(t_i)\right)$ is the deterministic function of $\mathbf{n}(t_i)$ that can be shown to be finite for all $\mathbf{n}(t_i)\in \Omega$.
Then
\begin{equation}\label{eq: current vs size}
\lim_{t\to\infty}	\frac{\mathbb{E}Y_t}{t}=\lim_{t\to\infty}\frac{1}{t}\mathbb{E}\sum_{i=1}^{\mathfrak{N}_t(p)} \mathbb{E} \left(S_i|\mathbf{n}(t_i);\{\mathfrak{N}_s(p),s\in [0,t]\}\right)=p \mathbb{E}_{st}   S,
\end{equation}
where in the right hand side there is an  expectation of the avalanche size $S$ over the stationary state. To arrive at the result we perform averaging in two steps. The internal expectations of avalanche sizes are conditioned on a particular realization of  the Poisson process, i.e. on the values of random  jump times $t_1,t_2,\dots$, as well as  on the particle configurations the avalanches start with, so that there is a bounded deterministic function of configurations under the sum, while the external expectation averages out these quantities.  
Using  the law of large numbers for the Poisson process 
\begin{equation}
\lim_{t\to\infty}\frac{\mathfrak{N}_t(p)}{t}=p\quad a.s. 
\end{equation} 
and boundedness of $\mathbb{E} \left(S_i|\mathbf{n}(t_i)\right)$ one can interchange the external expectation and the $t\to\infty$ limit and 
replace $\mathfrak{N}_t(p)$ in the upper summation limit by $pt$. Then using the ergodic theorem for positive-recurrent Markov chains we replace the time averaging over the sequence   $\{\mathbf{n}(t_i)\}_{i\in \mathbf{N}}$ by averaging over the stationary state.

One can see that up to the factor $p$ the mean current coincides with the stationary average avalanche size.  	What about the diffusion coefficient?  In terms of the avalanche sizes and the above Poisson process we have
\begin{equation}\label{key}
\lim_{t\to\infty}\frac{\langle Y_t^2\rangle_c}{t}=\lim_{t\to\infty}\frac{1}{t}\mathbb{E}\sum_{i=1}^{\mathfrak{N}_t(p)} \sum_{j=1}^{\mathfrak{N}_t(p)} \langle S_iS_j \rangle_c,
\end{equation}
where 	$\langle S_iS_j \rangle_c=\mathbb{E}S_iS_j -\mathbb{E}S_i\mathbb{E}S_j$ is the two-time covariance. The presence of terms with $i\neq j$ does not allow one to reduce the problem to averaging  over the stationary state. It would be so, if different avalanches  were not correlated,
i.e.  $\langle S_iS_j \rangle_c =\delta_{ij}\langle S_i^2 \rangle_c $, in which case the diffusion coefficient would be proportional to the stationary state variance of the  avalanche size,   
\begin{equation}\label{eq: Delta =S^2}
\Delta=p\langle S^2 \rangle_{st;c},
\end{equation}
just like the current was proportional to its mean.  
As we will see, at low densities, when the avalanches are finite, correlations play a significant role in changing the asymptotic behaviour of $\Delta$. At high densities the avalanches become large, going through the lattice many times. It is natural to expect that in this case they effectively reset the system so that it almost forgets its state before the avalanche started. In other words, the correlations in the system with large avalanches are expected to become subdominant, so that (\ref{eq: Delta =S^2}) would hold at least asymptotically.        

\subsection{Finite size  results}
Before going to the results for the cumulants of $Y_t$ we note that it can be readily seen from (\ref{eq: master free},\ref{eq: boundary_cond})
that the stationary measure of the AAP on a ring is a uniform measure, i.e. the stationary probability of a configuration $\mathbf{n}$ is 
\begin{equation}
P_{st}(\mathbf{n})=\frac{1}{Z(N,p)},
\label{P_st} 
\end{equation}
where
\begin{equation}\label{key}
\qquad Z(N,p) = C_N^p
\end{equation}
is the number of stable configurations. 
It also can be thought of as a simplest example of the factorized measure 
with the one site weights \begin{equation}
{P}_{st}(\mathbf{n})=Z^{-1}(N,p) \prod_{i=1}^N f(n_i)\delta_{||\mathbf{n}||,p}	
\end{equation} $f(n)= \delta_{n,1}+\delta_{n,0}$  conditioned to the fixed number of particles, eq.  (\ref{eq: n=p}). 
Then, it is not unexpected that the expectation of an observable $O(\mathbf{n})$ over the stationary state  can be represented as  a contour integral of the form   
\begin{equation}\label{st expect}
\mathbb{E}_{st} O(\mathbf{n})=\oint_{\Gamma_0} \hat{O}(z)D_{N,p}(z),\label{eq: intrep}
\end{equation}  
where the integration contour is a simple loop closed around $z=0$ and leaving all the other singularities outside, $\hat{O}(z)$ is the generating function corresponding to the observable $O(\mathbf{n})$ and following to 
\cite{TP2020}    we introduce  the \textit{normalized differential}
\begin{equation}\label{eq: D_{N,p}}
D_{N,p}(z) := \frac{dz}{2\pi\mathrm{i}} \frac{1}{Z(N,p)} \frac{F^N(z)}{z^{p+1}}
\end{equation}
expressed via the generating function of one-site stationary state weights 
\begin{eqnarray}
F(z) = \sum_{n=0}^\infty z^n f(n)= 1 + z.\label{eq: F(z)}
\end{eqnarray} 
In particular, this is the case for the average particle current, which indeed can be represented as the stationary state observable. Furthermore, as it will be shown below,  integrals of the same structure are the main ingredients of the formulas of the second scaled current cumulant as well as, conjecturally, of the higher cumulants, which are beyond the scope of the stationary state observables.     

Evaluating   derivative (\ref{eq: lambda'(0)}) of the largest eigenvalue of the deformed generator of AAP we obtain the following exact expression for the integral current
\begin{equation} \label{eq: j_Nresult_int_form}
J(N,p) =N\left( R j^R_N - L j^L_N\right),  
\end{equation}
where the right and left jump contributions are given by integrals similar to (\ref{eq: intrep}),
\begin{equation} \label{eq:j_R,j_L def}
j^R_N =  (1-q) \oint_{\Gamma_0} D_{N,p}(z) z g'(zq), \qquad j^L_N =  (1-q) \oint_{\Gamma_0}  D_{N,p}(z) z g'(z),
\end{equation}
and  the function $g(z)$  is defined as a series
\begin{equation} \label{eq: g_def}
g(z) = - \frac{1}{(1-q)} \sum_{k=1}^{\infty} \frac{(-z)^{k}}{[k]_q} = \sum_{i=0}^{\infty} \frac{q^{i}z}{1+q^{i}z}.
\end{equation}

Similarly, evaluating  (\ref{eq: lambda''(0)}) we obtain for the diffusion coefficient
\begin{equation} \label{eq: Delta_result_int_form}
\Delta =  R  \Delta^R - L \Delta^L,
\end{equation}
where both $\Delta^R$ and  $ \Delta^L$  are given by  
\begin{eqnarray} \label{lambda_2R} \nonumber
\Delta^I = \epsilon(I) p N j_N^I 
+2 N^2  \sum_{i=0}^{\infty} \oint\oint D_{N,p}(t) D_{N,p}(y) t    \frac{ a^I(y)}{t-q^i y}
\\
+2 N^2 \sum_{i=1}^{\infty} \oint\oint D_{N,p}(t) D_{N,p}(y) t   \frac{ a^I(q^i y)}{t-q^i y}
\end{eqnarray}
with  $I \in \{R,L\}$,   $\epsilon(R) = 1, \epsilon(L) = -1$  and
\begin{eqnarray}
a^R(y) = (1-q) yg'(qy) -\frac{yj^R_N}{\rho(1+y)}, \quad \label{eq: a^R,a^L}
a^L(y) = (1-q) yg'(y) -\frac{yj^L_N}{\rho(1+y)}.
\end{eqnarray}
Formulas (\ref{eq: j_Nresult_int_form}-\ref{eq: a^R,a^L}) state the first main result of the article.

These integral representations suit well for further asymptotic analysis. At the same time, unlike e.g. the integrals in similar formulas obtained for the q-boson ZRP in \cite{TP2020}, the ones obtained here can explicitly be evaluated to sums due to the simple form  (\ref{eq: F(z)}) of the function $F(z)$. In particular, for the average current we have  
\begin{equation} \label{eq: jN as a sum}
J(N,p) = N\frac{(1-q)}{C_N^p}\sum_{m=0}^{p-1} (m+1)\frac{(-1)^{m} C_N^{p-m-1}}{1-q^{m+1}}(Rq^m - L).
\end{equation}
Similarly, this can be done for the diffusion coefficient. However, the result looks a lot more cumbersome, and we leave its derivation as an exercise for the reader.

As we have already mentioned, the mean integral current can be obtained as an average over the stationary state. 
It is nothing but the average avalanche size, where the single-particle jumps are included and the left ones are counted with the minus sign  times the rate of their beginning. However, the use of the formula (\ref{eq: intrep}) still requires some effort to construct the function $\hat{O}(z)$ out of the non-local avalanche size observable. Alternatively, this can be done in the integral form by exploiting the connection of AAP with the discrete time q-Boson ZRP \cite{PM2006} or in the form of the sum by considering discrete time model having the AAP as a limit, as it was done in \cite{PPC2003} for the totally asymmetric version, $R=1, L=0$,  of AAP. We refer the reader to \ref{sec: stationary current} for the derivation of the formula (\ref{eq: jN as a sum})   using the arguments generalizing those of  \cite{PP2005}.

The formulas \eqref{eq: j_Nresult_int_form} and \eqref{eq: Delta_result_int_form} can be used for the calculation of the mean particle current and diffusion coefficient for the systems with a small number of particles. For, example, with one particle on the lattice, they  expectedly  produce  the $q$- and N-independent  result
\begin{equation}
J (N,1) = R-L, \qquad \Delta(N,1) = R+L.\nonumber
\end{equation}
In  the system with $p=2$ particles the resulting expressions are more complex
\begin{eqnarray}\nonumber
J(N,2) = \frac{2N}{N-1} \left( R-L -\frac{2(Rq-L)}{N(1-q)} \right),
\end{eqnarray}
\begin{eqnarray}\nonumber
\Delta(N,2) = 2 N \left(R j^R_{N,p=2}+L j^L_{N,p=2}\right) + \\ \nonumber
+\frac{4N}{3 \left(N-1\right)^2\left(1+q\right)\left(1-q^2\right)} 
\Big( R\left(\left(N-2\right) \left( -1 + 2 q - q^2 -3 q^3 \right) - 3 q^2 N \right) -\\ \nonumber \qquad \qquad \qquad \qquad \qquad \qquad \qquad - L\left( \left(N-2\right) \left(1 + 3q - q^3\right) + 3q N \right)  \Big).
\end{eqnarray}
having nontrivial dependence on $q$ and $N$. Though, the $q$-dependence disappears from the mean current in the infinite system size $N\to\infty,$
\begin{equation}
J(\infty,2) = 2(R-L), \\ \nonumber
\end{equation}
reproducing the result for two independent particles, it still survives in the diffusion coefficient
\begin{eqnarray} \nonumber
\Delta(\infty,2) = 4(R+L) - \frac{4 R (1 - 2 q + 4 q^2 + 3 q^3) + 4 L (-1 - 6 q + q^3)}{3 (1 - q^2) (1 + q)}.
\end{eqnarray}
In the   $q= 0$ limit this result reproduces $\Delta  = \frac{8}{3}\left(R+L\right)$ obtained  for the totally asymmetric  drop-push model, $L=0$,  in \cite{SRB1994} and for its partially asymmetric  generalization studied in  \cite{PP2005} with  $L\neq 0$. 

Of course, with more calculational effort both the integral current and diffusion coefficient can be given the form of rational functions of $q$ and $N$ for a few larger values of $p$. Of physical interest, however, is their behavior in systems of large size, which we describe in the next subsection. 

\subsection{Scaling limits}
Here we show  the asymptotic form our exact formulas take in the thermodynamic limit
\begin{equation}\label{eq: thermodynamic limit}
N,p\to\infty,\quad\rho=p/N=\mathrm{const}.
\end{equation}
The behaviour of  particle current 	in this limit has already been studied in \cite{PIPH2001,PPC2003_AAP,PPC2003}.
It was shown to have a transition at the critical density 
\begin{equation}\label{eq: crit density}
\rho_c=\frac{1}{1-q}.
\end{equation} 
Specifically, the current per site has different asymptotic bahaviour below, above and at the critical point,
\begin{eqnarray}\label{eq: current}
\!\!\!\!\!\!\!\!\!\!\!\!\!\!\!\!\!\!\!\!\!\!\!\!\!\!\!\!\!\!\!\!\!\!\!\!\!\!\!\!\!\!	j_N(\rho)=\left(1+O(N^{-1})\right) \times \left\{	\begin{array}{ll}	\frac{\rho(1-\rho)(R\rho_c+(1-\rho_c)L)}{(\rho-\rho_c)^2}+	j_\infty^{\mathrm{reg}}(\rho), & \rho<\rho_c, \vspace{2mm}	\\
N(R\rho_c+L(1-\rho_c)), &  \rho=\rho_c, \vspace{2mm}\\
N^{3/2} e^{Ns(\rho|\rho_c)} \frac{\sqrt{2 \pi \rho(1-\rho)} }{\rho_c(1-\rho_c)} (\rho-\rho_c)(\rho_cR+(1-\rho_c)L), &  \rho>\rho_c,	
\end{array}\right.\,
\end{eqnarray}
where 
\begin{equation}\label{eq: current reg}
j_\infty^{\mathrm{reg}}(\rho)=	\frac{\rho_cR+(1-\rho_c)L}{\rho_c(1-\rho_c)}\sum_{k=1}^\infty k\frac{\left[\frac{(\rho_c-1)^2}{\rho-1}\frac{\rho}{\rho_c^2}\right]^k}{1-\left[\frac{\rho_c-1}{\rho_c}\right]^k}-\frac{L\rho(1-\rho)}{\rho_c}
\end{equation}
is  a regular part of the sub-critical current  that remains finite  as the density  $\rho$ approaches its critical value $\rho_c$ from below and 
\begin{equation}\label{eq: s(rho|rho_c)}
s(\rho|\rho_c)=(1-\rho)\ln\left(\frac{1-\rho}{1-\rho_c}\right)+\rho\ln\left(\frac{\rho}{\rho_c}\right)
\end{equation}
is the so called relative entropy of the Bernoulli distribution with parameter $\rho$ relative to the one with parameter $\rho_c$. It is  non-negative approaching zero only at   $\rho=\rho_c$.   

One can see that the particle current per site $j_N(\rho)$ approaches a finite value in the thermodynamic limit in the sub-critical regime $\rho<\rho_c$. This value, however, diverges as $\rho\nearrow \rho_c$. The divergent term is explicitly shown in the first line of  (\ref{eq: current})  having a critical exponent  $\nu=2$ \cite{PIPH2001}. 

Exactly at the critical point, $\rho=\rho_c$, the current grows linearly with $N$. Being proportional to the average avalanche size the linear growth indicates that  the avalanche size distribution in the infinite system has a power-law tail. 

Finally in the super-critical regime, $\rho>\rho_c$, the current grows exponentially in $N$ with the exponent $Ns(\rho|\rho_c)$ times the power law prefactor. The appearance of the relative entropy in the exponent is common for the large deviation theory. For example, $s(\rho|\rho_c)$  is the large deviation rate function describing the probability for the frequency of heads and tails in tossing the unfair coin to be $\rho$ and $1-\rho$, when the heads and tails probabilities are $\rho_c$ and $1-\rho_c$ respectively \cite{Var}. In our case, the appearance of this function as an exponent of the average avalanche size compiles with the random walk picture of the avalanches described below. 

The $L=0$ case of  (\ref{eq: current}-\ref{eq: s(rho|rho_c)})  was  obtained before   in \cite{PIPH2001,PPC2003} from the analysis of the stationary state. The  formula for  arbitrary $R$ and $L$ was obtained  from the asymptotic analysis of the Bethe ansatz solution \cite{PIPH2001,PPC2003_AAP} only in the subcritical regime.

For the asymptotic behaviour of the diffusion coefficient we have 
\begin{eqnarray}\label{eq: Delta^j_N asymp}
\!\!\!\!\!\!\!\!\!\!\!\!\!\!\!\!\!\!\!\!	\Delta_N^j(\rho)&	\!\!=&\!\!\!\left(1+O(N^{-1/2})\right) \\ \nonumber &\times& \left\{	\begin{array}{ll}	N^{-1/2}\left(\frac{\sqrt{\pi}(R\rho_c+L(1-\rho_c))(\rho_c(1-\rho_c))^{3/2}(\rho_c^2-2\rho_c(1-\rho)-\rho)}{2(\rho-\rho_c)^4}   + \Delta_{\infty}^{reg}(\rho)\right), & \rho<\rho_c \vspace{2mm}	\\
N^{3/2}(R\rho_c+L(1-\rho_c)) \sqrt{\pi\rho_c(1-\rho_c)}, &  \rho=\rho_c \vspace{2mm}\\
N^{2} e^{2Ns(\rho|\rho_c)} 4\pi (\rho-\rho_c)(R\rho_c +L(1-\rho_c)) \frac{\rho(1-\rho) }{\rho_c(1-\rho_c)} , &  \rho>\rho_c	
\end{array}\right.\,
\end{eqnarray}
where in the sub-critical regime we again  explicitly show the part of the leading asymptotics that diverges when the density approaches its critical value, while its regular part is given in terms of a convergent series  
\begin{eqnarray}\label{eq: Delta^{reg}}
\!\!\!\!\!\!\!\!\!\!\!\!\!\!\!\!\!\!\!\!\!\!\!\!\!\!\!\!\!\!\!\!\!\!\!\!\!\!\!\!	\Delta_\infty^{\mathrm{reg}}(\rho)=	\frac{\sqrt{\pi}(R\rho_c +L(1-\rho_c))}{4\sqrt{\rho(1-\rho)}\rho_c(1-\rho_c)}\sum_{k=1}^\infty  \frac{\left[\frac{(\rho_c-1)^2}{\rho-1}\frac{\rho}{\rho_c^2}\right]^k }{1-\left[\frac{\rho_c-1}{\rho_c}\right]^k} \left(k^2 (1-2\rho) - k^3\right)-\frac{\sqrt{\pi}(\rho(1-\rho))^{3/2}}{4\rho_c } L.
\end{eqnarray}

As it was noticed in \cite{PPC2003_AAP}   the diffusion coefficient shows the typical KPZ behaviour  in the sub-critical regime. Specifically, let us consider  an interface  on the cylinder of   circumference $N$  with  the height function defined as a piece-wise constant function  $h(x,t),$   with jumps at  integer coordinates  equal to the occupation numbers at corresponding sites  and  helicoidal boundary conditions consistent with the average tilt $\rho$, 
\begin{equation}\label{eq: mapping}
n_i(t)=h(i,t)-h(i+0,t), \quad h(x+N,t)=h(x,t)-\rho N,\quad 
\end{equation}    
where $i\in \mathbb{Z}/N\mathbb{Z},\quad x\in \mathbb{R}.$ In the large time limit, $t\to \infty,$  the interface height dominanted by the position of the center of mass of the interface typically grows linearly in time,  $h(x,t)\simeq t j_N(\rho),$ with the mean velocity $j_N(\rho)$ that asymptotically approaches $j_\infty(\rho)$ as the system becomes large, $N\to\infty$. Its variance $\langle h^2(x,t)\rangle_c \simeq t\Delta^j_N(\rho)$ also  grows linearly with the rate equal to the per site diffusion coefficient. For the KPZ interface  the scaling picture developed in  \cite{KMH} suggests that the latter is expected 
to have the large $N$ asymptotics 
\begin{equation} \label{sc_beh_F_Lt}
\Delta^j_N(\rho)= c_0 A^{\frac{3}{2}} |\lambda| N^{-\frac{1}{2}}
\end{equation}
expressed in terms of two dimensionful invariants $A$ and $\lambda$, which can be constructed from the stationary state observables of the process in the traslation invariant infinite system and the universal constant $c_0=\sqrt{\pi}/4$  defined earlier from exact solutions of other models \cite{DEM1993}   (for details see similar discussion in \cite{TP2020}).
In our case of the factorized stationary state with Bernoulli one site marginals, $\mathbb{P}_{st}(n_i=1)=\rho$,  the invariants are   
\begin{eqnarray}
A = \langle n_i^2 \rangle_{st;c}=\rho (1-\rho),\quad
\lambda =\frac{d^2j_\infty(\rho)}{d\rho^2}.
\end{eqnarray}
Using the subcritical $j_\infty(\rho)$ from (\ref{eq: current},\ref{eq: current reg}) one can check that   (\ref{sc_beh_F_Lt}) indeed coincides with
the subcritical part of $\Delta^j_N(\rho)$ from (\ref{eq: Delta^j_N asymp},\ref{eq: Delta^{reg}}). It is also in agreement with 
the sub-critical large deviation function  found in \cite{PPC2003} from the asymptotic analysis of the Bethe ansatz solution. 
Note that the asymptotic method of \cite{PPC2003} per se was tightly related to the KPZ scaling and produced already an asymptotic form of the currently scaled cumulants.  
In contrast, starting with the exact formula here we are capable to obtain also the critical and supercritical asymptotics of the diffusion coefficient.  As one can see from (\ref{eq: Delta^j_N asymp})
they show the power law and exponential growth with $N$ respectively. The critical exponent  $3/2$ of the former defines the scale in which the crossover takes place. The description of the crossover is the subject of the next subsection. 	

\subsection{Crossover}
 Here we consider simultaneous limits
 \begin{equation}\label{eq: crossover scaling}
 N\to\infty \quad \mathrm{and} \quad \rho\to\rho_c 
 \end{equation}  
 such that 
\begin{equation}\label{eq: beta_def}
		\beta = \frac{\sqrt{N} (\rho_c-\rho) }{\sqrt{\rho_c(1-\rho_c)}}
\end{equation}
playing the role of crossover parameter remains finite. 
Then, the particle current is given by
	\begin{equation}\label{eq: j crossover}
		j_{N}(\rho) = N (R\rho_c+L(1-\rho_c))  {\mathcal F}(\beta) + O(N^\frac{1}{2}), 
	\end{equation}
	with the scaling function  defined by 
	\begin{equation} \label{j_N_sc_function}
		\mathcal{F}(\beta) = 1 - \sqrt{\frac{\pi}{2}}\beta \  {\rm erfc} \left(\frac{\beta}{\sqrt{2}}\right)e^{\frac{\beta^2}{2}}
	\end{equation}
	in terms of complementary error function
	${\rm erfc} (x) = \frac{2}{\sqrt{\pi}}\int_x^{+\infty} e^{-t^2} dt.$ This function was first obtained in \cite{PPC2003} for the totally asymmetric version of the process with $L=0$ from the analysis of the stationary state of the discretized AAP. The $O(N^{1/2})$ correction term in \eqref{eq: j crossover} explicitly given below  in Section \ref{subsec: crossover regime} disappears at $\beta = 0$, yielding a $O(1)$ correction.
	
\begin{figure}%
	\centering
	\subfloat[]{{\includegraphics[width=7cm]{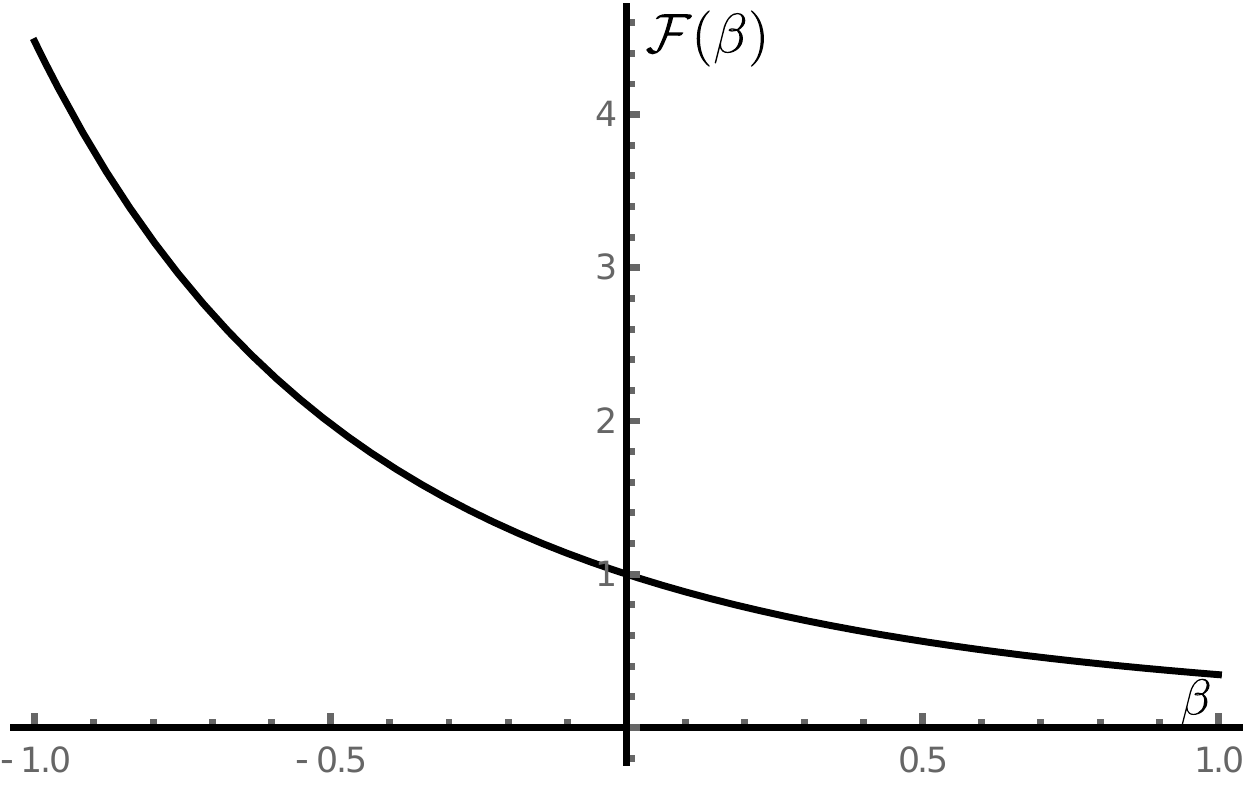}}} \quad
	\subfloat[]{{\includegraphics[width=7cm]{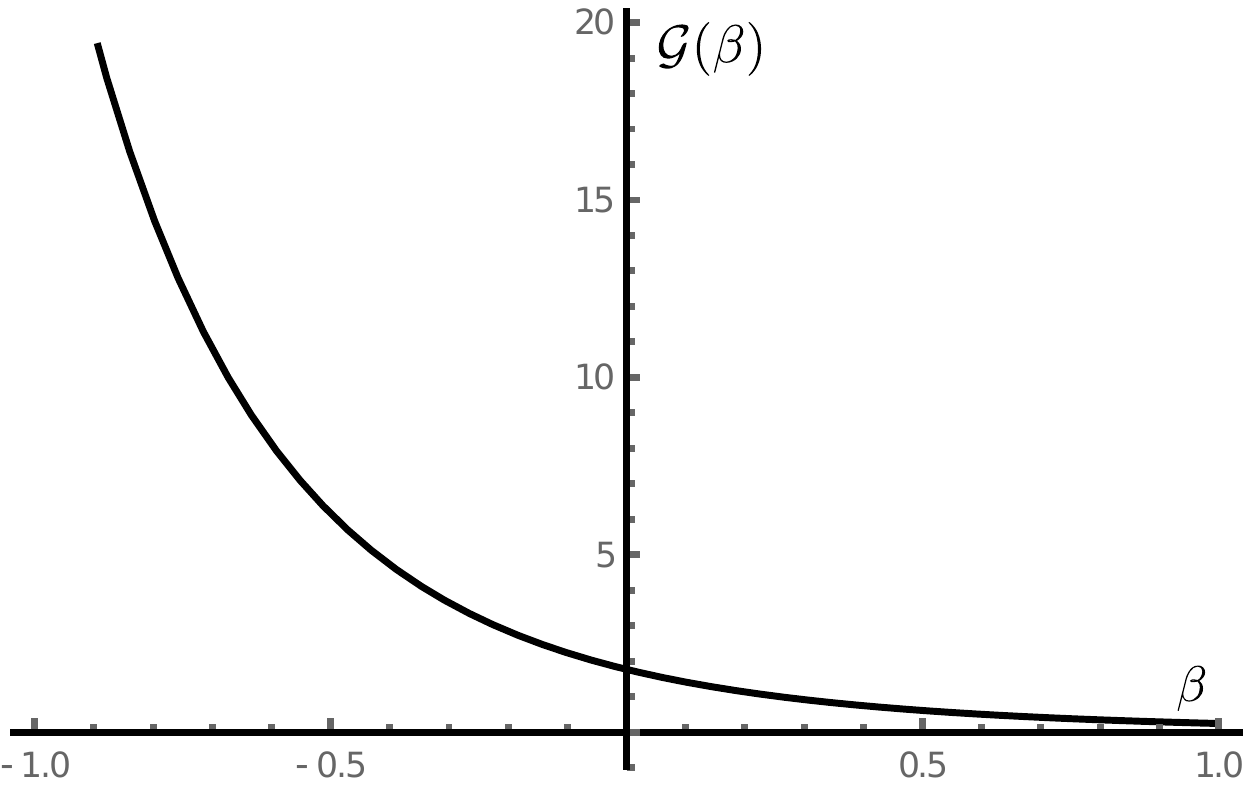} }}%
	\caption{Scaling functions (a) $\mathcal{F}(\beta)$ and (b) $\mathcal{G}(\beta)$.}%
	\label{fig: sc_fun}
\end{figure}

The function $\mathcal{F}(\beta)$ monotonously decreases  from infinity to zero as its argument runs over the real axis (see Fig. \ref{fig: sc_fun}(a)).
The  leading orders of its asymptotic expansions 
	\begin{eqnarray}
		\mathcal{F}(\beta) &=& \frac{1}{\beta^2} -\frac{3}{\beta^4}+ O(\beta^{-6}), \quad \beta \rightarrow +\infty,\\
		\mathcal{F}(\beta) &=& 1 -\sqrt{\frac{\pi}{2}}\beta + O(\beta^2), \quad \beta \rightarrow 0, \\
		\mathcal{F}(\beta) &=&  -\beta \sqrt{2\pi} e^{\frac{\beta^2}{2}} + O(\beta^{-2}), \quad \beta \rightarrow -\infty
	\end{eqnarray}
match with the divergent term of the sub-critical 
expression of $j_\infty(\rho)$,  its critical and supercritical bahaviours (\ref{eq: current}) respectively. 

The per site diffusion coefficient under the same scaling is  
	\begin{eqnarray} \label{eq: crossover_dif_coef}
		\Delta^j_{N}(\rho)=  N^{\frac{3}{2}} \left(R\rho_c+L(1-\rho_c)\right) \sqrt{ \rho_c (1-\rho_c)}  \ \mathcal{G(\beta)}  + O(N), 
	\end{eqnarray}
	where the scaling function
	\begin{equation}
	\label{eq: Sc_fun_dif_coef}
	\mathcal{G(\beta)} = \sqrt{\pi} (2{\mathcal F}(\sqrt{2} \beta) - {\mathcal F}\left( \beta \right))
	\end{equation}
	is defined in terms of ${\mathcal F}\left( \beta \right)$ from (\ref{j_N_sc_function}).
	This scaling function, Fig. \ref{fig: sc_fun}(b), also interpolates between the sub-critical, critical and super-critical regimes of the diffusion coefficient, and its asymptotic expansions 
	\begin{eqnarray}
		\mathcal{G}(\beta) =\frac{3 \sqrt{\pi}}{2 \beta^4} - \frac{45 \sqrt{\pi}}{4 \beta^6} + O(\beta^{-8}), \quad \beta \rightarrow +\infty,\\
		\mathcal{G}(\beta) =  \sqrt{\pi} + \frac{\pi}{\sqrt{2}} (1-2\sqrt{2}) \beta + O(\beta^2), \quad \beta \rightarrow 0,\\
		\mathcal{G}(\beta) = -4\pi\beta e^{\beta^2} - \sqrt{2}\pi\beta e^{\frac{\beta^2}{2}}  +O(\beta^{-4}), \quad \beta \rightarrow -\infty\label{eq: beta to infty}
	\end{eqnarray}
match with the leading behaviours of three regimes from (\ref{eq: Delta^j_N asymp}). 

\section{Crossover functions and first passage area  for the  Ornstein-Uhlenbeck    process \label{sec: Crossover and first passage}}
Returning to the discussion at the end of subsection \ref{sec: Model observables} we recall that up to the simple  factor the mean current is  the average stationary avalanche size, while the diffusion coefficient is expected to behave asymptotically as the variance of the avalanche size when typical avalanches are large. Thus, it would be interesting to compare these quantities. Instead of studying the exact statistics of avalanches, we perform this comparison in the scaling limit, in which the evolution of the number of particles in the active site within an avalanche can be described by a stochastic PDE. Remarkably, this is exactly the scaling limit corresponding to the crossover between the sub- and super-critical regimes described above.  

Consider first the totally asymmetric version of AAP with $L=0,R=1$. It was pointed in \cite{PPC2003}  that the number of particles jumping from the active site during an avalanche performs a biased random walk with the coordinate dependent bias. Specifically the number  $\chi(k)$  of particles going out of  the active site at the avalanche step number $k$  can either increase or decrease by one or remain the same on the next step  with probabilities $\mathbb{P}(\chi(k+1)=b|\chi(k)=a)$ vanishing unless  $a-b=\pm1,0$. An avalanche starts with $\chi(0)=1$ and ends at the first return to the origin $\chi(T)=0$, where  $T=\min\left(k>0:\chi(k)=0\right) $  be the number of the step,  at which $\chi(k)$ hits  the origin for the first time. Then,
the avalanche size $S$ is given by the sum
\begin{equation}
S=\sum_{k=1}^T \chi(k).
\end{equation}
The stationary state random walk transition probabilities are   
\begin{eqnarray}\label{eq: RW}
\!\!\!\!\!\!\!\!\!\!\!\!\!\!\!\!\!\!\!\!\!\!\mathbb{P}(\chi(k+1)=b|\chi(k)=a)\simeq \left\{	
\begin{array}{ll}	\left(1-\left(\rho-\frac{a}{N}\right)\right)(1-\mu_a), & b=a-1, 	\\
\left(1-\left(\rho-\frac{a}{N}\right)\right)\mu_a+	\left(\rho-\frac{a}{N}\right)(1-\mu_a), &  b=a, \\
\left(\rho-\frac{a}{N}\right)\mu_a, & b=a+1.	
\end{array}\right.
\end{eqnarray}
for $a,b>1$ and we note that the limit $\mu_a\to (1-\rho_c)$ is approached exponentially quickly as $a\to\infty$. 

Let us introduce a rescaled process
\begin{equation}
X_t^N=(\rho_c(1-\rho_c)N)^{-1/2}\chi([t N]).
\end{equation}
It is not difficult to show that  in the limit $N\to\infty$ it converges in law to  a version of the Ornstein-Uhlenbeck  \cite{OU,Doob} process $X_t$, known as  Vasicek model first  introduced in \cite{V}  for financial applications, which  satisfies the following stochastic PDE
\begin{equation}
dX_t=-(\beta+X_t)dt+\sqrt{2}dW_t,\label{eq: O-U}
\end{equation}	
where $W_t$ is the standard Wiener process and $\beta$ is the  parameter introduced in (\ref{eq: beta_def}) that is   supposed to remain finite in the limit (\ref{eq: crossover scaling}). Let us also suppose that $X_t$ starts at 
\begin{equation}
X_0=\alpha>0
\end{equation} with probability one and is stopped at the time \begin{equation}
\tau=\inf(t\in\mathbb{R}_{\geq 0}:X_t=0),
\end{equation}	
when $X_t$  reaches the origin. Then the rescaled avalanche size will correspond to the area $\mathcal{A}(\alpha)$ under the trajectory of $X_t$ until the first passage of the origin,
\begin{equation}
\mathcal{A}(\alpha)=\int_0^\tau X_t dt.
\end{equation}

A vast literature on the first passage problems  exists  motivated by 
both the development  of probability theory and stochastic processes as well as  by plenty applications to natural sciences, qualitative finance e.t.c., see  \cite{Redner} for review. Many explicit formulas have been  obtained for the processes related to the  Brownian motion, which can be found e.g. in \cite{BorodinSalminen} together with recipes for deriving similar formulas that are not listed. Among them, for example, the Laplace transform of the probability density of the  time of first exit beyond the constant boundaries  of the Ornstein-Uhlenbeck process, which was first obtained  in \cite{BellmanHarris1951} back in fifties of the last century. The  general formulation  for    arbitrary    homogeneous strongly continuous Markov processes
shortly followed \cite{DarlingSiegert1953} as well as its extention to a similar problem  for the generating functions of additive functionals on these processes \cite{DarlingSiegert1956}. In both cases the problem under suitable conditions on the transition probabilities is reduced to a simple second order ODE. The Laplace transforms or generating functions of interest  given by  solutions of the ODE are represented in terms of   special functions and can not be explicitly inverted except for a few simplest cases. Instead, one can obtain  the moments  of corresponding distributions treating the Laplace transforms as the moment generating functions.  In particular it was shown in \cite{DarlingSiegert1953} that the moments of the first exit time satisfy simple contiguous differential relations, which can often be explicitly integrated.

Following these ideas a few first  moments  of $\mathcal{A}(\alpha)$ and $\tau$ were recently derived   in \cite{KM2021} for the Ornstein-Uhlenbeck process, i.e. the particular case of (\ref{eq: O-U}) with $\beta=0$. Adapting these arguments for general $\beta$ we find that  the generating function
\begin{equation}
\tilde{P}(s|\alpha)=\mathbb{E}e^{-s\mathcal{A}(\alpha)}
\end{equation}
satisfy the following ODE
\begin{equation}
\left[\frac{d^2}{d\alpha^2}-(\beta+\alpha)\frac{d}{d\alpha}-s\alpha\right]\tilde{P}(s|\alpha)=0
\end{equation}
subject to boundary conditions $\tilde{P}(s|0)=1$ and  $\lim_{\alpha\to\infty}\tilde{P}(s|\alpha)=0$.
Differentiating  this equation in $s$ and setting $s=0$, we obtain a relation between the contiguous moments of the area 
\begin{equation}
\mathcal{A}_n(\alpha)=\mathbb{E}\mathcal{A}(\alpha)^n=(-1)^n\left.\frac{d^n \tilde{P}(s|\alpha)}{ds^n}\right|_{s=0}
\end{equation} 
having a form 
\begin{equation}
\left[\frac{d^2}{d\alpha^2}-(\beta+\alpha)\frac{d}{d\alpha}\right]\mathcal{A}_n(\alpha)=-n\alpha\mathcal{A}_{n-1}(\alpha);\quad \mathcal{A}_0(\alpha)\equiv 1.
\end{equation}
subject to initial conditions $\mathcal{A}_n(0)=0$. It is solved by the recursion
\begin{equation}
\mathcal{A}_n(\alpha)=n\int_0^\alpha e^{\frac{1}{2}(z+\beta)^2} \int_z^\infty z'e^{-\frac{1}{2}(z'+\beta)^2}\mathcal{A}_{n-1}(z')dz'dz
\end{equation}
that yeilds the following expressions for the first and the second moments
\begin{eqnarray}
\mathcal{A}_1(\alpha)&=&\int_0^\alpha e^{\frac{1}{2}(z+\beta)^2} \int_z^\infty z'e^{-\frac{1}{2}(z'+\beta)^2}dz'dz\\
\mathcal{A}_2(\alpha)&=&2\int_0^\alpha dz_1 e^{\frac{1}{2}(z_1+\beta)^2} \int_{z_1}^\infty dz_2 z_2 e^{-\frac{1}{2}(z_2+\beta)^2}\\
&\times&\int_0^{z_2} dz_3  e^{\frac{1}{2}(z_3+\beta)^2} \int_{z_3}^\infty dz_4 z_4 e^{-\frac{1}{2}(z_4+\beta)^2}.
\end{eqnarray}
To return back to the moments of the avalanche size we should rescale the moments of the area as    $\mathcal{A}_n(\alpha)\to \left(N^{3/2} \sqrt{\rho_c(1-\rho_c)}\right)^n \mathcal{A}_n(\alpha)$ and set $\alpha=1/\sqrt{\rho_c(1-\rho_c)N}$. Then, to the leading order in $1/\sqrt{N}$ we obtain 
\begin{eqnarray}
\mathbb{E}_{st}S&\simeq&N\mathcal{F}(\beta), \label{eq: ES}\\
\mathbb{E}_{st}S^2&\simeq&N^{5/2}\sqrt{\rho_c(1-\rho_c)}\mathcal{J}(\beta)\label{eq: ES^2},
\end{eqnarray}
where $\mathcal{J}(\beta)$ is yet another scaling function
\begin{equation}
\mathcal{J}(\beta)=\left(\frac{2(1-\mathcal{F}(\beta))}{\beta}-4\beta\mathcal{F}(\beta)+e^{\frac{\beta^2}{2}}\pi\beta^2\int_\beta^\infty e^{\frac{x^2}{2}}\left(\mathrm{erfc}\left(\frac{x}{\sqrt{2}}\right)\right)^2dx\right).
\end{equation}
Since the square of (\ref{eq: ES}) is of smaller order than (\ref{eq: ES^2}), the latter also yields the cumulant  $\langle S^2\rangle_{st;c}$ to the leading order.
To generalize these formulas to arbitrary $L$ and $R$, we should correct them by explicitly taking the left Poissonian jumps into account. To this end, we should multiply both formulas by  probability $(qL+R)$ of the birth of an avalanche from two neighboring occupied sites. These are the avalanches that bring the leading order contribution to the mean avalanche size, while the Poissonian jumps, which  lead to one step avalanches, are subdominant.  As a result, (\ref{eq: ES}) together with (\ref{eq: current vs size}) exactly reproduces (\ref{eq: j crossover}). 

The function $\mathcal{J}(\beta)$ is to be compared with $\mathcal{G}(\beta)$. Its asymptotic behaviour, corresponding to 
the three regims, is as follows
\begin{eqnarray}\label{eq: J(beta)}
\mathcal{J}(\beta)&=& 	
 10\beta^{-5}-130\beta^{-7}+ O(\beta^{-9}),  \ \beta \rightarrow + \infty,	\\
\mathcal{J}(\beta)&=& \sqrt{2\pi} - 6 \beta + O(\beta^2),   \ \beta \rightarrow 0, \\
\mathcal{J}(\beta)&=& -4 \pi \beta e^{\beta^2} (1 + \beta^{-1}+ O(\beta^{-2})),   \ \beta \rightarrow - \infty	.	
\end{eqnarray}

\begin{figure}%
		\subfloat[]{{\includegraphics[width=7cm]{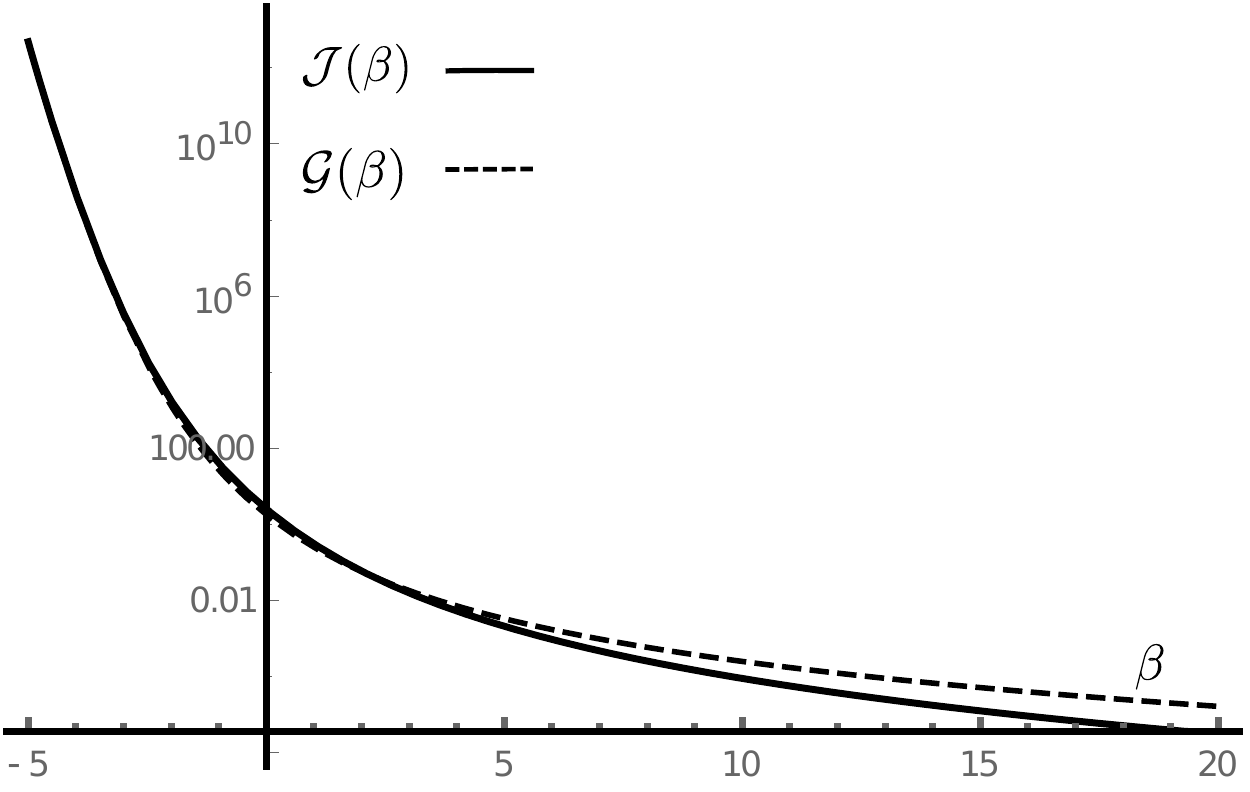}}}\quad
	\subfloat[ ]{{\includegraphics[width=7cm]{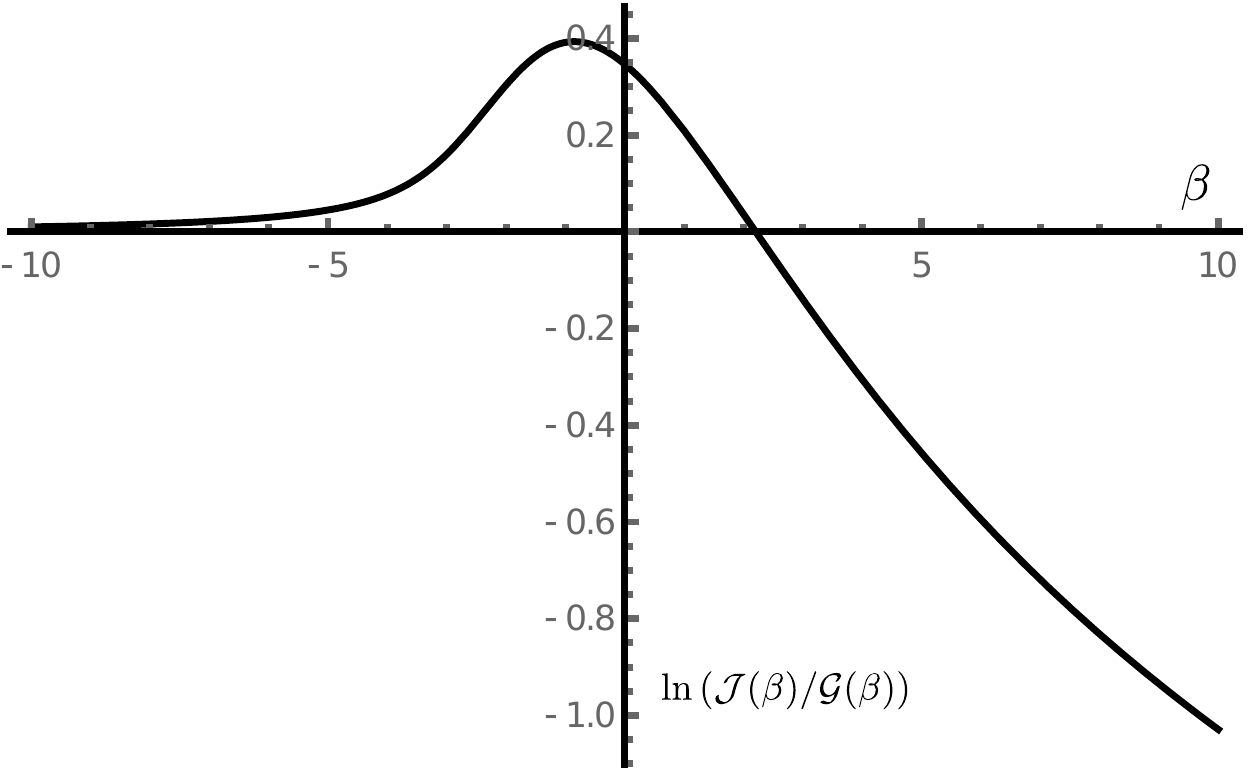} }}%
	\caption{Comparison of scaling functions $\mathcal{J}(\beta)$ and $\mathcal{G}(\beta)$.  (a) Plots of  $\mathcal{J}(\beta)$ and $\mathcal{G}(\beta)$} in log scale. The functions are asymptotically similar as $\beta\to-\infty$, while the latter dominates the former in the opposite limit. (b) Plot of the logarithmic ratio of the functions. It vanishes  in the limit $\beta\to-\infty$ and diverges to negative infinity in the limit $\beta\to\infty$.
	\label{fig: G-J}
\end{figure}

One can see that $\mathcal{J}(\beta)$ agrees with $\mathcal{G}(\beta)$
  to the leading order, when $\beta\to-\infty$, having the corrections of order of $O(1/\beta)$ times the leading terms instead of exponentially smaller corrections in (\ref{eq: beta to infty}). Thus, (\ref{eq: ES^2}) indeed agrees with (\ref{eq: Delta =S^2}) only asymptotically, as it was expected from discussion in subsection \ref{sec: Model observables}, see Fig. \ref{fig: G-J}a. 
On the other hand, the greater  the parameter $\beta$ is, the less is the ratio  $\mathcal{J}(\beta)/\mathcal{G}(\beta)$ vanishing as $\beta\to\infty$, Fig. \ref{fig: G-J}b.  The fact that the scaling function $\mathcal{G}(\beta)$ describing the scaled variance of the current dominates $\mathcal{J}(\beta)$ responsible for the variance of the avalanche size in the subcritical region is an indication of positive correlations between the subsequent avalanches dominating  in the KPZ regime. 

\section{The exact expressions for  scaled current cumulants} \label{sec: TQ}

Having introduced the exact integral representations for the mean particle current and diffusion coefficient we now discuss in detail how these are derived using Bethe ansatz and T-Q equation method. 	In this section, we calculate the first two derivatives of the largest eigenvalue of the deformed operator $\mathcal{L_{\gamma}}$.

\subsection{Reformulation of the Bethe equations}
The operator $\mathcal{L_{\gamma}}$ for AAP is diagonalized by Bethe ansatz 
\cite{PPC2003_AAP}. Therefore, we start off with the Bethe ansatz equations (BAE)  
\begin{equation} \label{BAE}
e^{-\gamma N}\Big(\frac{1+x_k}{1+qx_k}\Big)^N = (-1)^{p-1}\prod_{j=1}^p \frac{x_k-q x_j}{x_j-q x_k},\quad k=1,\dots,p,
\end{equation} 
which define complex numbers $x_{1},\dots,x_{p}$ to be substituted to the formula 
\begin{equation}
\lambda(\gamma)= \sum_{k=1}^p \left(R \frac{1+x_k}{1+qx_k} + L \frac{1 + q x_k}{1+x_k}\right)-p.\label{eq: lambda}
\end{equation}
of eigenvalue of the operator $\mathcal{L_{\gamma}}$. We need to identify the solution of (\ref{BAE}) corresponding to the largest eigenvalue.  This solution is defined by condition $x_i \to 0, \ i=1,\dots,p$ as $\gamma\to 0 $ that ensures  $\lambda(0)=0$. Also identity
\begin{equation}\label{TI}
e^{-\gamma p} \prod_{k=1}^p \frac{1+x_k}{1+q x_k} = 1,
\end{equation}
following from the translation invariance of the corresponding eigenstate  holds for this solution, when  $\gamma$ is small enough.

To reformulate Bethe equations into a functional equation for  polynomials in one variable we define
a  degree $p$ polynomial 
\begin{equation}\label{eq: Q}
Q(x) = \prod_{i=1}^p (x-x_i)
\end{equation}
in an auxiliary variable $x$ with $p$ roots at the roots  $x_1, \dots, x_p$ of BAE. Then, we write the condition of divisibility of another polynomial,   obtained from (\ref{BAE}) and having  zeroes at  $x_1,\dots,x_p$,  by  $Q(x)$  as the functional relation
\begin{equation} \label{TQ}
T(x) Q(x) = e^{-\gamma N} \left(1+x\right)^N Q(qx)+\left(1+qx\right)^N q^p Q\left(x/q\right), 
\end{equation}
between $Q(x)$ and yet another polynomial  $T(x)$ of degree $N$.	
The formulas (\ref{eq: lambda},\ref{TI}) can also be rewritten in terms of $Q(x)$	
\begin{equation} \label{lambda_via_Q}
\lambda(\gamma)= Rq^{-2}(1-q) \frac{Q'(x)}{Q(x)}\Big|_{x= -1/q} - L(1-q) \frac{Q'(x)}{Q(x)}\Big|_{x=-1} + p\left(Rq^{-1}+Lq-1 \right),
\end{equation}
\begin{equation} \label{ini_cond}
e^{-\gamma p} Q(-1)= q^p Q\left(-1/q\right).
\end{equation}
Following to \cite{PM2008} we are going to solve T-Q relation \eqref{TQ} perturbatively in powers
of $\gamma$ in the vicinity of $\gamma=0$. Let us consider  an expansion of the polynomials $T(x), Q(x)$ and the  eigenvalue $\lambda(\gamma)$ in powers of~$\gamma$ 
\begin{eqnarray*}
	Q(x) & = & Q_{0}(x)+\gamma Q_{1}(x)+\gamma^{2}Q_{2}(x)+\dots\text{,}\\
	T(x) & = & T_{0}(x)+\gamma T_{1}(x)+\gamma^{2}T_{2}(x)+\dots,\\
	\lambda(\gamma) & = & \gamma\lambda_{1}+\gamma^{2}\lambda_{2}+\dots, 
\end{eqnarray*}
where  from $\lambda(0)=0$ we know that $\lambda_0=0$. The relation \eqref{TQ} is equivalent to the linear system of equations for the
polynomials $T_{k}(x)$ and $Q_{k}(x)$. This system can be solved order by order with initial conditions
\begin{equation}
Q_{0}(x)=x^{p},\ \ \ T_{0}(x)=(1+x)^{N} q^p + (1+qx)^N.\label{0 order}
\end{equation}
In this paper we are interested in the first two coefficients of the eigenvalue which are related to the scaled cumulants
$$
\text{\ensuremath{\lambda_{1}=J, \quad}\ensuremath{\lambda_{2}=\frac{\Delta}{2}}}.
$$
To obtain the exact expressions for them we solve T-Q relation in the first and second orders.

\subsection{First order calculation}
In the first order TQ-equation \eqref{TQ}  becomes
\begin{equation}
\!\!\!\!\!\!\!\!\!\!	T_{0}(x)Q_{1}(x)+T_{1}(x)Q_{0}(x)=(1+x)^N (Q_{1}(qx) - NQ_{0}(qx))+(1+qx)^{N}q^{p}Q_{1}(x/q). \,\,\,\,\,\,\,\,\,\,\label{TQ 1 order}
\end{equation}
To solve this equation in polynomials we use the observation that the degree of polynomial $Q_1(x)$ is at most $p-1$. Therefore, it is enough to solve this equation $\mathrm{modulo}\,\,x^{p}$.
We define  
\begin{equation}
B_{1}(x)=q^{p}Q_{1}(x/q)-Q_{1}(x)\label{def_of_B_1}
\end{equation}
to present the equation \eqref{TQ 1 order} as follows 
\begin{equation}\label{eq: diff eq uniform}
(1+qx)^{N}B_{1}(x) = (1+x)^N B_{1}(qx)\,\,\mathrm{mod}\,\,x^{p}.\label{B_1 eq}
\end{equation}
The solution  $\tilde{B}_{1}(x) $ defined up to a multiplicative constant $\tilde{B}_{1}(0)$ is
\begin{equation}
\tilde{B}_{1}(x)  =  \tilde{B}_{1}(0) (1+x)^{N}.
\end{equation}
As  the desired polynomial $B_{1}(x) = \sum_{i=0}^{p-1}b_{i}x^{i}$ is at most degree $p-1$, we use an integral representation to extract the necessary terms. The constant $\tilde{B}_{1}(0)$ is defined from the initial condition \eqref{ini_cond},  which gives $B_1(-1) = (-1)^{p-1} p$   in first order in $\gamma$. The resulting representation for $B_1(x)$ is
\begin{equation}
B_1(x) = \frac{N x^p}{Z(N,p)}  \oint\limits_{|z|<|x|} \frac{F(z)^N}{z^p} \frac{1}{x-z} \frac{dz}{2\pi\mathrm{i}}.
\end{equation}
Here the integration contour is a simple anticlockwise loop around
the origin $z=0$, which must be the only singularity inside the contour. To turn back to the polynomial $Q_{1}(x)=\sum_{i=0}^{p-1}q_{i}x^{i}$ we use \eqref{def_of_B_1} to find the relation between   the coefficients of $Q_1$ and $B_1$
\begin{equation}
b_{i}=(q^{p-i}-1)q_{i}\label{bviaq}.
\end{equation}
Now we can represent  the relation between the polynomials $Q_{1}(x)$ and $B_{1}(x)$ in the integral form 
\begin{eqnarray}\label{eq:Q_1}
Q_{1}(x) & = & -x^{p}\oint\frac{B_{1}(z)}{z^{p+1}}\sum_{i=1}^{\infty}\frac{(z/x)^{i}}{1-q^{i}}\frac{dz}{2\pi\mathrm{i}}.
\end{eqnarray}
Then, we note  that a substitution of $\tilde{B}_{1}(x)$ instead of $B_{1}(x)$ does not change the result of integration, as only  $p$ terms of the sum in the integrand contribute to the result. The integral representation of  $Q_{1}(x)$ in terms of function $g(z)$ defined by \eqref{eq: g_def} is 
\begin{eqnarray}\label{Q_1(x)_int_form}
Q_{1}(x) & = & \frac{N x^{p}}{ Z(N,p)} \oint\frac{F(z)^N}{z^{p+1}} g\left(-\frac{z}{x}\right) \frac{dz}{2\pi\mathrm{i}}.
\end{eqnarray}
The first derivative of the eigenvalue $\lambda(\gamma)$ in terms of $Q_0(x)$ and $Q_1(x)$ follows from expansion of  \eqref{lambda_via_Q} to the first order  of  $\gamma$ 
\begin{eqnarray*}
	\!\!\!\!\!\!\!\!\!\!\!\!\!\!\!\!\!\!\!\!\!\!\!\!\!\!\!\!\!\!		\lambda_{1} = (1-q)\left( Rq^{-2}
	\frac{ Q_1'(x) Q_0(x) - Q_1(x)Q_0'(x)}{Q_0^2(x)}\Big|_{x=-1/q} 
	- L\frac{ Q_1'(x) Q_0(x) - Q_1(x)Q_0'(x)}{Q_0^2(x)}\Big|_{x=-1}\right)
\end{eqnarray*}
where $f'(z) = \partial_{z} f(z)$.
Substituting  the integral form of $Q_1(x)$ we obtain
\begin{eqnarray}\label{lambda_1_result}
\lambda_{1} = \frac{(1-q)N}{Z(N,p)} \oint \frac{F(z)^N}{z^p} \Big[R g'(qz) - L g'(z) \Big].
\end{eqnarray}
This coincides with the result obtained from averaging over the stationary state \eqref{J_int_form}.

\subsection{Second order calculation}
Similarly to the first order, the second order TQ-equation \eqref{TQ} can be rewritten in terms of polynomial
\begin{equation}
B_{2}(x)=q^{p}Q_{2}(x/q)-Q_{2}(x)
\end{equation}
and has the following form
\begin{eqnarray}\nonumber
(1+x)^N B_2(qx) = (1+qx)^N B_2(x) - T_1(x)Q_1(x) -N (1+x)^N Q_1(qx)+\\
+ x^p\left( \frac{q^p N^2}{2} (1+x)^N - T_2(x)\right)
\end{eqnarray}
with the initial conditions obtained from \eqref{ini_cond}
\begin{eqnarray} 
B_2(-1) = \frac{(-1)^p p^2}{2} - p \ Q_1(-1) \label{B_2(1)}.
\end{eqnarray}
Repeating the same reasoning, we present the second order relation so that the polynomial in the left hand side has the power at least $p-1$. For these first $p-1$ terms  we solve equation modulo $x^p$ \begin{eqnarray} 
B_2(qx) - \frac{(1+qx)^N}{(1+x)^N} B_2(x)= - (1+x)^{-N}T_1(x)Q_1(x)-N Q_1(qx) 
\quad\mathrm{mod}\,\,x^{p}, \label{B_2 eq} 
\end{eqnarray}
where we notice that $T_1(x)$ is a polynomial of degree not exceeding $N+p-1$ known from the first order calculations \eqref{TQ 1 order}
\begin{eqnarray}
T_{1}(x)=-N q^{p}(1+x)^N + x^{-p}\left((1+qx)^N B_{1}(x)-(1+x)^{N}B_{1}(qx)\right).
\end{eqnarray}
We introduce $q$-difference operator
\begin{equation}
D_{q}a(x)=\frac{a(qx)-a(x)}{(q-1)x}
\end{equation}
to represent the equation for  $B_2(x)$ in the form of the first order linear inhomogeneous $q$-difference equation with non-constant coefficients
\begin{equation}
D_{q}B_{2}(x)=\frac{f(x)}{(q-1)x}+\frac{(1+qx)^{N}(1+x)^{-N}-1}{(q-1)x}B_{2}(x)\mathrm{\,\,mod}\,\,x^{p},\label{eq for B_2}
\end{equation}
where $f(x)$ stands  for the right hand side of  \eqref{B_2 eq}.
To solve the
$q$-difference equation we first find a solution to the corresponding homogeneous equation which is once again $S F(z)^N$ up to some constant $S$. The constant variation method yields  the general solution of (\ref{eq for B_2})
\begin{eqnarray}
\tilde{B}_{2}(x)=S F(x)^{N}-\sum_{i=0}^{\infty}\frac{f(q^{i}x)}{F^{N}\left(q^{i+1}x\right)}F(x)^{N} ,
\end{eqnarray}
where $S$ is the constant of integration defined by \eqref{B_2(1)}. Coming back to the polynomial $Q_{2}$ we use the same relation \eqref{bviaq} between the coefficients of polynomials $Q_2(x)$ and $B_2(x)$. The result is the integral representation for $Q_{2}(x)$
\begin{equation}
Q_{2}(x)=x^{p}\oint\frac{F^{N}(z)}{z^{p+1}}\left(S-\sum_{i=0}^{\infty}\frac{f(q^{i}z)}{F^{N}\left(q^{i+1}z\right)}\right)g\left(-z/x\right)\frac{dz}{2\pi\mathrm{i}}.
\end{equation}
The constant $S$ is found from \eqref{B_2(1)}
\begin{equation}
S = \frac{1}{C_{N-1}^{p-1}}\Big((-1)^{p} Q_1(-1)p -\frac{p^2}{2} + \oint \frac{F^{N-1}(z)}{z^{p}} \sum_{i=0}^{\infty}\frac{f(q^{i}z)}{F^{N}\left(q^{i+1}z\right)}\frac{dz}{2\pi\mathrm{i}}\Big).
\end{equation}
The resulting expression for $Q_2(x)$ is
\begin{eqnarray}\label{eq: Q_2(x)}
Q_2(x) = \left((-1)^p Q_1(-1) -\frac{p}{2}\right)Q_1(x) -
\oint \frac{F^{N}(z)}{z^{p}} \sum_{i=0}^{\infty}\frac{f(q^{i}z)}{F^{N}\left(q^{i+1}z\right)}
\phi(z,x)\frac{dz}{2\pi\mathrm{i}},
\end{eqnarray}
where we introduce function
\begin{equation}
\phi(z,x) = \frac{x^{p}}{z} g\left(-z/x\right) -\frac{Q_1(x)}{p(1+z)} .
\end{equation}
The next step is to simplify this result by substituting explicit expression  of the polynomial   $f(x)$. Using (\ref{eq: diff eq uniform}) we rewrite $f(x)$ as follows
\begin{eqnarray}\label{eq: f(x) represent}
\!\!\!\!\!\!\!\!\!\!\!\!\!\!\!\!\!\!\!\!\!\!\!\!\!\!\!\!\!\!\!\!\!\!\!\!\!\!\!\!	f(x) = N B_{1}(qx) - \frac{Q_1(x)}{x^p}\left(B_1^{\geq p}(qx) - \frac{(1+qx)^N}{(1+x)^N} B_1^{\geq p}(x)\right)\, \mathrm{mod} \, x^p\\
\!\!\!\!\!\!\!\!\!\!\!\!\!\!\!\!\!\!\!\!\!\!\!\!	=  - \left[\frac{Q_1(x)}{x^p}B_1^{\geq p}(qx)-\frac{q^p N B_{1}(qx)}{q^p-1}\right] + \left[\frac{Q_1(x)}{x^p}\frac{(1+qx)^N}{(1+x)^N} B_1^{\geq p}(x)-\frac{ N B_{1}(qx)}{q^p-1}\right] \, \mathrm{mod} \, x^p\nonumber
\end{eqnarray}
in terms of the complementary  polynomial $B^{\geq p}_1(x)=		\tilde{B}_{1}(x)-B_1(x)$ obtained by removing the terms  of degrees less than $p$ in  $	\tilde{B}_{1}(x)$, which in turn has the following 
integral representation
\begin{equation}
B_{1}^{\geq p}(z) = \frac{ N}{C_N^p} z^p \oint\limits_{|z|<|t|} \frac{dt}{2\pi\mathrm{i}} \frac{F^{N}(t)}{t^{p}(t-z)}.  
\end{equation}
We note that  convergence of the infinite sum under the integral in (\ref{eq: Q_2(x)}) is guaranteed by the absence of a  free term in  $f(x)$, which  can be checked by a direct calculation, $f(0) = N Q_1(0)(q^p-1) -Q_1(0)(q^p b_p   - b_p) = 0$ using  $b_p = N$ and by the fact that  the denominator  approaches the limit $F^N(q^{i+1}z)\to1$  as $i\to\infty$. 	In the second line of (\ref{eq: f(x) represent})
we represent $f(x)$ as the sum of two polynomials  (expressions in square brackets mod $p$) having no the free term each. Therefore, the part of the sum in (\ref{eq: Q_2(x)}) with each of these terms individually  is absolutely convergent, and one can rearrange the summands by grouping the terms containing $B^{\geq p}_1(xq^i)$ with the same $i$ as it was done in \cite{TP2020}. Since the resulting series is absolutely convergent we can interchange the integration and summation. Finally, noticing that the integrals of $F^N(z)B_1(q^{i}z)/(z^pF^N(q^{i}z))\phi(x,z)$ vanish we arrive at	 
\begin{eqnarray}\nonumber
Q_2(x) &=& \left((-1)^p Q_1(-1) -\frac{p}{2}\right) Q_1(x)-\oint
\frac{dz}{2\pi\mathrm{i}}
\frac{\phi(z,x)}{z^{2p}} B_1^{\geq p}(z) Q_1(z)\\ \label{Q_2_via_B_1>p}
&-&\frac{N}{C_N^p} \sum_{i=0}^{\infty} \oint \frac{dz}{2\pi\mathrm{i}} \frac{F^{N}(z)}{z^p} \oint
\frac{dy}{2\pi\mathrm{i}} \frac{F^{N}(y)}{y^{p}} \frac{B_1^{\geq p}(q^{i+1}z)}{F^{N}\left(q^{i+1}z\right)} \frac{ \phi(z,x)}{(q^{i+1}z-y)}.
\end{eqnarray}
Note that after evaluation of vanishing integrals the sum and integrals in \eqref{Q_2_via_B_1>p}   can not be interchanged back anymore. 

In the next step we substitute the integral representation for both $B_{1}^{\geq p}(x)$ and $Q_{1}(x)$ and obtain
\begin{eqnarray}\nonumber
\!\!\!\!\!\!	\!\!\!\!\!\!\!\!\!\!\!		\!\!\!\!\!\!\!\!\!\!\!		\!\!\!\!\!\!\!\!\!\!\!		Q_2(x) &\!\!\!\!\!\!	=&
\left((-1)^p Q_1(-1) -\frac{p}{2}\right) Q_1(x)+\\
\!\!\!\!\!\!\!\!\!\!\!		&\!\!\!\!\!\!	-&		\frac{N^2}{(C_N^p)^2}\Big[\oint\oint \frac{dy}{2\pi\mathrm{i}} \frac{F^{N}(y)}{y^{p+1}}\frac{dt}{2\pi\mathrm{i}} \frac{F^{N}(t)}{t^{p}} \oint\limits_{|y|<|z|<|t|}\frac{dz}{2\pi\mathrm{i}}
\frac{\phi(z,x)}{z(t-z)} g\left(-y/z\right)\\
\!\!\!\!\!\!\!\!\!\!\!		&\!\!\!\!\!\!	+&		 \sum_{i=0}^{\infty} \oint\oint\frac{dy}{2\pi\mathrm{i}} \frac{F^{N}(y)}{y^{p}} \frac{dt}{2\pi\mathrm{i}} \frac{F^{N}(t)}{t^{p}}
\oint\limits_{|y|<|q^{i+1}z|<|t|}\frac{dz}{2\pi\mathrm{i}} \frac{F^{N}(z)}{F^N(q^{i+1}z)} \frac{\phi(z,x)}{(t-q^{i+1}z)^N} \frac{q^{(i+1)p}}{(y - zq^{i+1})} \Big]. \nonumber
\end{eqnarray}
Triple integrals can be reduced to double integrals by integrating over the variable $z$ by counting the residues inside the contours. 
The first term in the square brackets has  infinitely many  poles  of the function $g\left(-y/z\right)$ at $q^{i}y,\quad i=0,1,\dots$. 
The $i$-th summand  of the sum in the second term has the only contributing pole $z=y q^{-i-1}$. Thus, the double integral representation for $Q_2(x)$ is 
\begin{eqnarray}\nonumber
Q_2(x) &=& 
\left((-1)^p Q_1(-1) -\frac{p}{2}\right) Q_1(x)\\ \nonumber &+&
\frac{N^2}{ Z(N,p)^2}\Big[ \sum_{i=0}^{\infty}\oint \oint\frac{dt}{2\pi\mathrm{i}} \frac{F^{N}(t)}{t^{p}}
\frac{dy}{2\pi\mathrm{i}} \frac{F^{N}(y)}{y^{p}} \frac{q^i \phi(q^i y, x)}{(t-q^i y)} \\
&+&\sum_{i=0}^{\infty} \oint\oint\frac{dt}{2\pi\mathrm{i}} \frac{F^{N}(t)}{t^{p}} \frac{dy}{2\pi\mathrm{i}} \frac{F^{N}(y)}{y^{p}} 
\frac{\phi(y, x)}{(t-q^{i+1}y)} \Big].
\end{eqnarray}
In terms of the  integrals with normalized differentials (\ref{eq: D_{N,p}}) it is given by
\begin{eqnarray}\nonumber
Q_2(x)& =& 
\left((-1)^p Q_1(-1) -\frac{p}{2}\right) Q_1(x)
+N^2  \oint \oint D_{N,p}(t) D_{N,p}(y)
t y \frac{\phi(y,x)}{t-y}\\
&+&N^2  \sum_{i=1}^{\infty} \oint \oint D_{N,p}(t) D_{N,p}(y)
t y \frac{q^i \phi(q^i y, x) + \phi(y,x)}{(t-q^i y)}. \label{Q_2_resum}
\end{eqnarray}
Finally, for the exact  expression of the second coefficient $\lambda_2$ of the eigenvalue \eqref{lambda_via_Q} we collect the terms with $\gamma^2$ obtaining
\begin{eqnarray} \nonumber
\lambda_2  =  R(1-q)q^{-2} m (-1/q) 
- L(1-q) m(-1),
\end{eqnarray}
where 
\begin{equation}
m(x) = \left(\frac{Q_2'(x)}{Q_0(x)} -\frac{Q_1'(x) Q_1(x)} {Q_0^2(x)}-\frac{Q_0'(x) Q_2(x)} {Q_0^2(x)} + \frac{Q_0'(x) Q_{1}^{2}(x)} {Q_0^3(x)}\right).
\end{equation}
The coefficients of  $R$ and $L$ are defined by the same function $m(x)$ given in terms of already known polynomials $Q_0(x)$, $Q_1(x)$, $Q_2(x)$ and calculated at points $-1/q$ and $-1$, respectively. The explicit form of the  function $m(x)$ obtained  from substitution of \eqref{Q_2_resum} is
\begin{eqnarray} \nonumber
m(x) &=& x^{-p} \Big((-1)^p Q_1(-1) -\frac{p}{2} - \frac{Q_1(x)}{x^p}\Big) \Big(Q_1'(x) -\frac{p Q_1(x)}{x} \Big)  \\ \nonumber
&+&N^2  \oint\oint D_{N,p}(t) D_{N,p}(y) t y  \frac{A(y,x)}{t-y}\\ 
&+&N^2  \sum_{i=1}^{\infty} \oint\oint D_{N,p}(t) D_{N,p}(y) t y  \frac{q^i A(q^i y,x) + A(y,x)}{t-q^i y},
\end{eqnarray}
where
\begin{equation}
\!\!\!\!\!\!\!\!\!\!\!\!\!\!\!	A(y, x) =x^{-p}\frac{d\phi(y,x)}{dx} -p x^{-p-1}\phi(y,x)=x^{-2} g'(-y/x) - \frac{Q_1'(x) -p x^{-1}Q_1(x)}{p x^p (1+y)}.
\end{equation}
For further convenience we introduce notations  $\lambda_2^R = (1-q)q^{-2} m \left(-q^{-1}\right)$, $\lambda_2^L = (1-q) m(-1)$ writing down the second coefficient of the eigenvalue as follows
\begin{equation} \label{Answer}
\lambda_2 =  R  \lambda_2^R - L \lambda_2^L.
\end{equation}
After substituting  $x = -q^{-1}$ and $x=-1$ into $m(x)$ and using  the second order initial condition \eqref{B_2(1)} we obtain
\begin{eqnarray} \nonumber
\lambda_2^R = \frac{p N j_R}{2} 
+ N^2 (1-q) q^{-2} \Big(\oint\oint D_{N,p}(t) D_{N,p}(y) t y  \frac{A(y,-q^{-1})}{t-y}  \\
\qquad  \qquad  + \sum_{i=1}^{\infty} \oint\oint D_{N,p}(t) D_{N,p}(y) t y \frac{q^i A(q^i y,-q^{-1}) + A(y,-q^{-1})}{t-q^i y}\Big),  \\ 
\nonumber
\lambda_2^L = -\frac{p N j_L}{2} 
+ N^2 (1-q) \Big(\oint\oint D_{N,p}(t) D_{N,p}(y) t y  \frac{A(y,-1)}{t-y} \\
\qquad  \qquad  + \sum_{i=1}^{\infty} \oint\oint D_{N,p}(t) D_{N,p}(y) t y \frac{q^i A(q^i y,-1) + A(y,-1)}{t-q^i y}\Big).
\end{eqnarray}
Finally introducing notations 
\begin{eqnarray}\label{eq: a^L(z),a^R(z) def.}
a^L(z) =z(1-q)A(z,-1),\\
a^R(t) = tq^{-2}(1-q)A(t,-q^{-1})
\end{eqnarray}	
we arrive at the result (\ref{eq: Delta_result_int_form}-\ref{eq: a^R,a^L}).

\section{Asymptotic analysis} \label{sec: Asymptotic analysis}

The purpose of is this section is the asymptotic analysis of the exact expressions for the integrated current and the  diffusion coefficient in the
thermodynamic limit (\ref{eq: thermodynamic limit}).
The main part is  evaluation of  integrals of the form 
\begin{equation}
\oint D_{N,p}(t) b(z) = \frac{1}{Z(N,p)}\oint \frac{dz}{2\pi\mathrm{i} z}e^{Nh(z)}b(z)\label{def_I}
\end{equation}
and its two-dimensional analogues. Here, 
\begin{equation}
h(z)=N^{-1}\ln \left(F(z)^N/z^p\right)=\ln (1+z)-\rho\ln(z)
\end{equation}
and	$b(z)$ is a function analytic in some vicinity of the origin $z=0$. 

\begin{figure}
	\centering{}\includegraphics[width=0.9
	\linewidth]{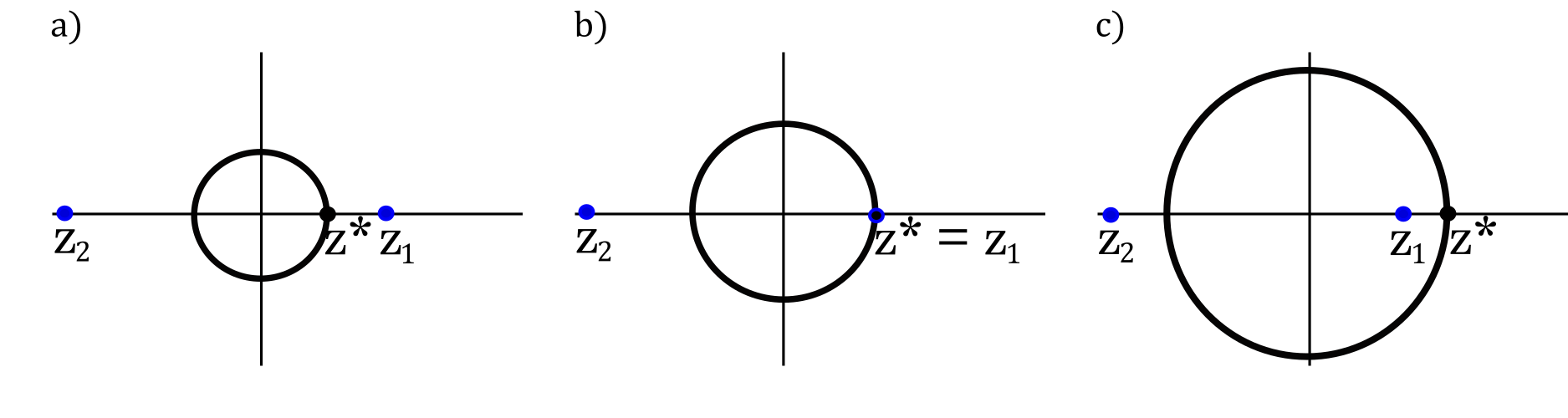}
	\caption{Steep descent contours for  a) the subcritical density where all the poles $z_i~=~-q^{-i}, ~i~=~1,2, \dots $  are outside the contour b) critical density when the saddle point $z^*$ coincides with the pole $z_1$ and c) supercritical density where one or more poles get inside the contour}
	\label{fig: contours}
\end{figure}

The function $h(z)$ has a single  critical point being a unique solution of equation 
\begin{equation}\label{eq: saddle_point eq}
h'(z^{*})=0
\end{equation}
that yields
\begin{equation} \label{saddle_point}
z^* = \frac{\rho}{1-\rho}.
\end{equation}
Then, we should deform the integration contour to the one   passing through   $z^*$, such that  the real part of $h(z)$ decreases monotonously on it away from the critical point. Such a contour will be referred to as steep descent in contrast to the steepest descent one, where also the imaginary part of  $h(z)$ is constant. One possible choice is the circle $z=z^*e^{\mathrm{i}\varphi}, \phi\in(-\pi,\pi),$
\begin{equation}\label{key}
\frac{d\Re h(z^*e^{\mathrm{i}\varphi})}{d\varphi}=-\frac{z^*\sin\varphi}{1+2z^*\cos{\varphi}+(z^*)^2}	\lessgtr 0,\quad \nonumber \mathrm{for}\quad 	\varphi \gtrless0, 
\end{equation}
respectively. 	Then the  standard saddle point estimate    gives in two leading orders in  $N$
\begin{equation}
\oint D_{N,p}(t)b(t)=b_{0}+\frac{1}{2N}\left(\frac{h_{3}b_{1}}{h_{2}^{2}}-\frac{b_{2}}{|h_{2}|}\right)+O(N^{-2}),
\label{eq:int D asymp}
\end{equation}
where $b_{k}=(z\partial_{z})^{k}b(z)|_{z=z^{*}}$ and $h_{k}=(z\partial_{z})^{k}h(z)|_{z=z^{*}}$, unless the contour being deformed 
passes through the singularities of the integrand.  If  poles (other than that in  $z=0$) turn out to be inside the contour, their contribution should be extracted from the saddle point contribution and may dominate the latter. In our case of the integrals in formulas of  both  current \eqref{eq: j_Nresult_int_form} and  diffusion coefficient \eqref{eq: Delta_result_int_form} the integrand  has poles of first and second orders at points $z_j = -q^{-j}, j = 1,2,\dots$ independent of the density. On the other hand, the higher is the density, the bigger is the value of $z^*$ being the radius of the saddle point contour. Therefore, being outside the contour at small densities, more and more poles enter the contour, when the density increases (see Fig. \ref{fig: contours}).  

\subsection*{$\rho<\rho_c$}

The poles are outside of the contour,  when $z^*<z_1$ , i.e.  $\rho<\rho_c$. In this case, applying (\ref{eq:int D asymp}) to (\ref{eq: j_Nresult_int_form},\ref{eq:j_R,j_L def})  we obtain for the current
\begin{equation}\label{eq: j_N result}
j_N(\rho)=	-(1-q) \sum_{k=1}^{\infty} \frac{k(-z^*)^k (Rq^{k-1}-L)}{1-q^{k}} \Big(1-\frac{k(2 \rho+k-1)}{2N \rho(1-\rho)}\Big) + O(N^{-2}) .
\end{equation}	
This formula obtained by direct substitution of formula (\ref{eq: g_def})  for the function $g(z)$  is in  fact valid in the 	narrower domain of the series convergence $|z^*|<1$. To extend it to $|z^*|<z_1$ one had to use  the re-summation of the  series representing  $g(z)$, which explicitly separates the  terms with poles at $z=-1$ and $z=z_1=-1/q$. Note that  the former is not a pole of the whole integrand, since it is also a zero of  $F^N(z)$. This re-summation yields the   analytic continuation 
of (\ref{eq: j_N result})  to the domain $0<z^*<z_1$, of which $O(1)$ part is given in (\ref{eq: current},\ref{eq: current reg}).  

A little more care is necessary to deal with the formulas (\ref{eq: Delta_result_int_form}-\ref{eq: a^R,a^L}) of diffusion coefficients. They are given in terms of series of double integral generalizations of (\ref{def_I}). Therefore, the saddle point calculations are performed in the same way in each of the integration variable $y$ and $t$ using the same saddle point and the steepest descent contour, except for the terms, where the integrand has a pole at $y=t$ connecting the two variables.

To get rid of this  term the integrand is represented as a half-sum of the symmetric and anti-symmetric in $t$ and $y$ terms. Then, the symmetric part will be regular at $t=y$ admitting a standard application of the saddle point method, while an integral of the anti-symmetric part in one of the variables, say $y$, can be reduced to that around the simple  pole $y=t$ and evaluated, resulting in 
\begin{eqnarray}\label{Pulling_contours}
\!\!\!\!\!\!\!\!\!\!\!\!\!\!\!\!	\oint\oint\limits_{|y|<|t|} & D_{N,p}(t)D_{N,p}(y)\frac{t a^I(y)}{t-y}=\\
& =  \frac{1}{2}\Big(\oint\oint D_{N,p}(t)D_{N,p}(y)\frac{ta^I(y)-ya^I(t)}{t-y}+ \frac{Z(2N,2p)}{Z(N,p)^2} \oint D_{2N,2p}(t)a^I(t)\Big).\nonumber 
\end{eqnarray}
Here $I=R,L$ like in (\ref{eq: Delta_result_int_form}-\ref{eq: a^R,a^L}) . Then, the two parts $\Delta^R$ and $\Delta^L$ of  (\ref{eq: Delta_result_int_form})  are represented by
\begin{eqnarray}\label{eq: Delta L} 
\!\!\!\!\!\!\!\!\!\!\!\!\!\!\!\!\!\!\!\!\!\!\!\!\!\!\!\!\!\!\!\!\!\!\!\!\Delta^I &= &N^2\left[\epsilon(I) \rho j_N^I +  \frac{Z(2N,2p)}{Z(N,p)^2}\oint D_{2N,2p}(t)  a^I(t)\right.\\ \nonumber
\!\!\!\!\!\!\!\!\!\!\!\!\!\!\!\!\!\!\!\!\!\!\!\!\!\!\!\!\!\!\!\!\!\!\!\!&-&\left. \oint\oint D_{N,p}(t) D_{N,p}(y)  \frac{ya^I(t)-ta^I(y)}{t-y}+2\sum_{i=1}^\infty \oint\oint D_{N,p}(t) D_{N,p}(y) t\frac{ a^I(q^iy) + a^I(y)}{t-q^{i}y} \right].
\end{eqnarray}
To estimate the leading asymptotics of the expression in square brackets we first recall that the integrals (\ref{eq:int D asymp})  with normalized differentials of functions analytic inside the contour are at most $O(1)$. One can also check that the following integral vanishes identically
\begin{equation}
\oint  D_{N,p}(y)a^I(y)=0. 
\end{equation}
This in particular suggests that 
\begin{equation}
a^I(z^*)=O(1/N).
\end{equation}
(One should keep in mind that $a^I(z^*)$ still has a  dependence on $N$ in the finite size corrections that come from  the $N$-dependence of $j_N(\rho)$.)
Therefore, the first integral associated with doubled $N$ and $p$ with the same $\rho$  is $O(1/N)$. It, however,  comes with the coefficient given by the ratio of partition functions that enhances the contribution of this term. The coefficient  is $O(\sqrt{N})$, so that the resulting order of magnitude is $O(1/\sqrt{N})$.   

Then, we argue that the $O(1)$ contribution from the other tree terms vanishes. To show this, we first 
note that the  integrand  in the second integral have the following  expansion  around the double saddle point $x=t=z^*$
\begin{equation}\label{eq: derivative expansion}
\frac{ya^I(t)-ta^I(y)}{t-y}=(z^*)^2\frac{d}{dz^*}\left(\frac{a^I(z^*)}{z^*}\right)+O(t-y), 
\end{equation}
so that the  leading $O(1)$ of the integral  is simply the constant part of (\ref{eq: derivative expansion}) and corrections are $O(1/N)$.

Next, using the following relations for the derivatives of function $g(z)$,
\begin{eqnarray}
zg'(z) = g(z)-g^2(z) - 2 \sum_{i=1}^{\infty}\frac{q^{i}}{1-q^{i}}\left(g(z)-q^{-i}g(q^{i} z ) \right), \label{eq:g' relation}\\
zg''(z) = -2 g(z)g'(z)-2\sum_{i=1}^{\infty}\frac{q^{i}}{1-q^{i}} \left(g'(z)-g'(q^{i} z)\right), \label{eq:g'' relation}
\end{eqnarray}
and equation (\ref{eq: saddle_point eq}) for the saddle point we find 	
\begin{eqnarray}
\frac{d}{dz^*}\left(\frac{a^I(z^*)}{z^*}\right) - 2\sum_{i=1}^{\infty} \frac{ a^I(q^i z^*)}{(z^*)^2(1-q^i)} -\epsilon(I)\frac{\rho j_N^R}{z^{*2}}= O(1/N),
\end{eqnarray} 
which proves that   all the $O(1)$ terms inside the square brackets vanish.  Finally, the result coming from the $O(1/N)$  part 
of the first integral calculated according to (\ref{eq:int D asymp}) is
\begin{equation} \label{Delta_L_via_A_and_h}
\Delta^I= \frac{N^{3/2}\sqrt{\pi}}{4 h_{2}^{3/2}}\left(a_{1}^I h_{3}-a_{2}^I|h_{2}|\right)+O(N),
\end{equation}
where $a_{k}^I = (z \partial_z)^k \big( a^I(z)\big)|_{z=z^*}$.
This justifies the $\rho<\rho_c$ part of (\ref{eq: Delta_result_int_form}-\ref{eq: a^R,a^L}).

Similarly to \cite{TP2020} we also note that	the first integral can exactly be evaluated as the difference of currents in systems of sizes $2N$ and $N$ at the same density
\begin{eqnarray}
\oint D_{2N,2p}(t) a^I(t) & = & j_{2N}^I(\rho) -j_N^I(\rho).\label{eq: j_2N-j_N}
\end{eqnarray}
Taking into account that 
\begin{equation}
\frac{Z(2N,2p)}{Z(N,p)^2}=\sqrt{2\pi N |h_2|}+O(1/\sqrt{N}),
\end{equation}
we find that the diffusion coefficient can be expressed in terms of the universal  finite size correction  $b=\lim_{N\to\infty}N(j_N(\rho)-j_{\infty}(\rho))$ to the particle current,
\begin{equation}
\Delta=-N^{3/2}b\sqrt{\pi  |h_2|/2}(1+O(1/\sqrt{N}).
\end{equation}
This is consistent with (\ref{sc_beh_F_Lt}) due to relation $b=-A\lambda/2$ between  the dimensionful invariants and the universal finite size correction first observed in \cite{KM1990} from studies of the KPZ equation. 

\subsection*{$\rho>\rho_c$}

\begin{figure}
	\centering{}\includegraphics[width=0.6
	\linewidth]{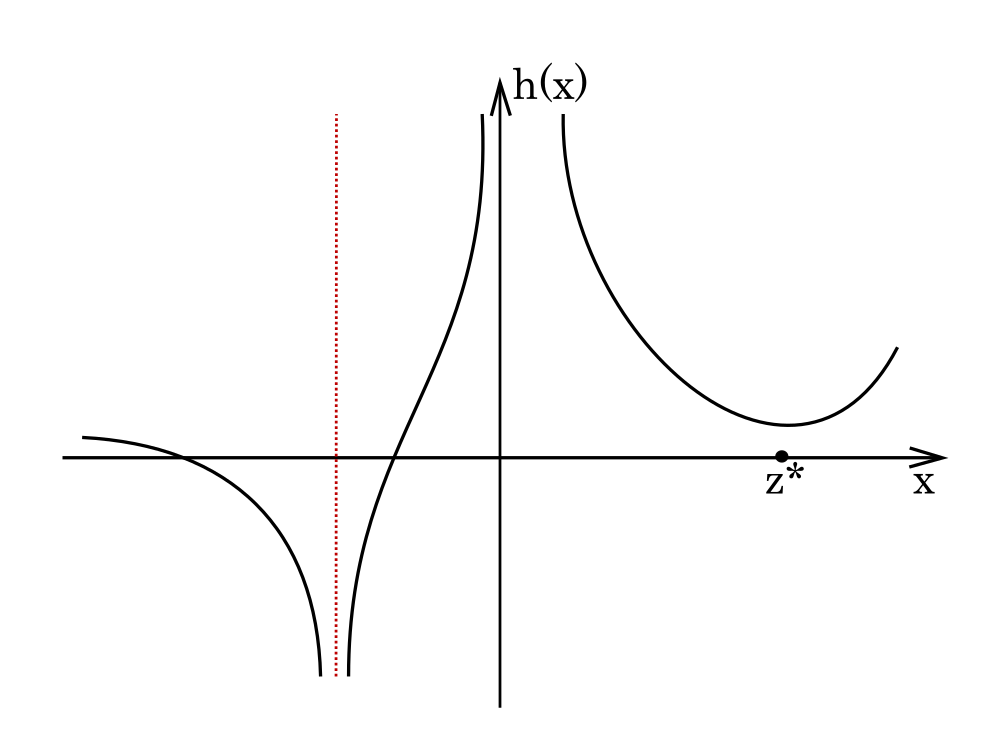}
	\caption{The sketch of the function $h(x)$ for the real valued variable $x$.}
	\label{h_function}
\end{figure}

When $z^*>|z_i|=|q^{-i}|$ for some $i\geq 1$ the poles $z=z_1,\dots,z_i$ are inside the contour and their contribution has to be extracted from the saddle point estimate. Asymptotically the contribution from the pole $z=z_i$  to (\ref{def_I}) has an exponential order   $O\left(e^{N  \Re h(z_i)}\right)$, possibly  with some power-law prefactor. Then, since
\begin{equation}\label{eq: h-ineq}
h(z_1)>h(z^*), \quad 	\mathrm{and} \quad \Re h(z_1)>\Re h(z_i), \ i\geq 2,
\end{equation}
we argue that the contribution of the pole at $z=z_1$  dominates both the other poles and the saddle point.
The inequalities (\ref{eq: h-ineq}) follow from the fact that $ h(z)$ is real and positive at the positive part of the real axis having a minimum at  $z=z^*$ (see Fig. \ref{h_function}), and for the negative poles we use $\Re h(z)\leq \Re h(|z|)$.    

To be specific, let us rewrite the exact expression  (\ref{eq: j_Nresult_int_form},\ref{eq:j_R,j_L def})   for the particle current per site separating the term with     the pole $z_1$
\begin{eqnarray} \label{eq: pole_decomp}
j_N(\rho) &=& (1-q)\Bigg[(R-Lq)\oint D_{N,p} (z)   \frac{z}{(1+qz)^2}  \\
&+&L\oint D_{N,p}(z)  \frac{z}{(1+z)^2}  +(R-Lq)\sum_{i=2}^{\infty}\oint D_{N,p}(z)   \frac{z q^{i-1}}{(1+q^i z)^2}\Bigg].\nonumber
\end{eqnarray} 
The second order pole $z_1$ of \eqref{eq: pole_decomp} brings the greatest exponential growth dominating both the saddle point and other poles' contributions resulting in
\begin{equation} \nonumber
j_N(\rho) = - \frac{(1-q)(R-Lq)}{Z(N,p)} \underset{z = z_1}{Res} \left( \frac{e^{N \ln h(z)}}{(1+qz)^2} \right)\left( 1+ O(e^{-cN})\right)
\end{equation}
for some $c>0$. Calculation of this residue and asymptotic representation of $Z(N,p) = e^{N h(z^*)}/\sqrt{2\pi N |h_2|}(1+O(1/N))$ justifies the third line of \eqref{eq: current}. 

Derivation of the supercritical asymptotics of the diffusion coefficient follows the same line. We first note that in  \eqref{eq: Delta L}   the terms of the largest order in $N$ are those, where the contribution comes from the pole $z_1$. This is the case  for the single integral, while among the double integrals we need to collect those, in which the poles $z_1$ in both variables are present. Also, the increase of the order of the pole by one brings the factor of  $N$ to the pre-exponential factor that comes from the differentiation of the exponent in $e^{Nh(z)}$.  With these comments in hand, we collect the highest order terms in  \eqref{eq: Delta L}

\begin{eqnarray}
\!\!\!\!\!\!\!\!\!\!\!\!\!\!\!\!\!\!\!\!\!\!\!\!\!\!\Delta&  = & N^2(R-Lq)\left[\frac{Z(2N,2p)}{Z(N,p)^2}\oint D_{2N,2p}(t) t \frac{(1-q)q}{(1+qt)^2} \right. \\ \nonumber
\!\!\!\!\!\!\!\!\!\!\!\!\!\!\!\!\!\!\!\!\!\!\!\!\!\!&+&\left. 2 \oint\oint D_{N,p}(t) D_{N,p}(y)yt \left(\frac{(1-q)q^2}{(1+qt)^2(1+qy)}-\frac{q j_N^L}{\rho(1+qy)(t-qy)}\right) \right]\left(1 + O(e^{-cN})\right)
\end{eqnarray}
for some $c>0$, which   after calculating the residues and substitution of the asymptotic expansions of the partition functions result in the third line of (\ref{eq: Delta^j_N asymp}).

\subsection*{$\rho=\rho_c$}
In this case, the pole  $z_1$ exactly coincides with the saddle point. This is a subtle case, as the saddle point method should be modified to cope with the singularity on the contour integration. In the next subsection we will do this modification to obtain the uniform asymptotics of the current and diffusion coefficient in the crossover regime. It turns out, however,  that the dominating term of particle current can be evaluated exactly \cite{PP2005}.

To see this, we note that two last terms in  (\ref{eq: pole_decomp}) are $O(1)$ computed as usual in the standard saddle point approximation.  
It follows then that 
\begin{equation} \label{eq: current_at_crit_density}
j_{N}(\rho_c) = (1-q) \frac{(R-Lq)}{Z(N,p)} I_{N,p,2}  + O(1),
\end{equation}
where we introduce the notation 
\begin{equation}\label{eq: I_{N,p,k}}
I_{N,p,k} = \oint \frac{(1+z)^N}{z^p}  \frac{1}{(1+qz)^k} \frac{dz}{2\pi\mathrm{i}}.
\end{equation}
Though $I_{N,p,k}$ is evaluated in terms of the hypergeometric function for general $N,p,k$ and $q$, for $k=2$ it simplifies drastically at the critical point. To see this we first do the integration by parts to show that   
\begin{equation}
I_{N,p,1}  = -\frac{1-q}{N} I_{N,p,2} + \frac{p}{N} I_{N+1,p+1,1}.
\end{equation}
From this identity  one can express $I_{N,p,2}$ as a sum of the two integrals, which being added up are reduced to a single integral evaluated to the  binomial coefficient using the fact that  $\rho=\rho_c$
\begin{equation}
I_{N,p,2}\Big|_{p/N = 1/(1-q)} = \frac{p}{(1-q)} I_{N,p+1,0} = \frac{p}{(1-q)}C_N^p.
\end{equation}
The substitution of the last expression into \eqref{eq: current_at_crit_density} gives
\begin{equation} \label{eq: j_cr}
j_{N}(\rho_c)=N(R\rho_c+L(1-\rho_c))  + O(1).
\end{equation}
This reproduces the second line of \eqref{eq: current}.

\subsection{Crossover regime} \label{subsec: crossover regime}
The above exact calculation based on a magic simplification of the integrand at the critical point seems not generalizable for the diffusion coefficient. However,  we can use a modification of the saddle point method for producing the uniform asymptotic estimates in the case when a pole merges with the saddle point \cite{Temme}. Specifically, let us consider the scaling limit (\ref{eq: crossover scaling},\ref{eq: beta_def}). In the vicinity of the critical density the estimate  (\ref{eq: current_at_crit_density}) still holds.

To evaluate   the integral $I_{N,p,2}$ from (\ref{eq: I_{N,p,k}})   asymptotically we first  make a variable change 
\begin{equation}
z=z^* e^{\mathrm{i}\varphi}
\end{equation}
going to go to integration in the  variable  $\varphi\in (-\pi,\pi]+\mathrm{i}a$ having   a small constant imaginary  part $a>0$ that ensures the pole $z=-1/q$   being outside of the integration contour in the original $z$-plane  or below the integration contour in the $\varphi$-plane, where we fix  $z^*=\rho_c/(1-\rho_c)=-1/q$.
For the contour still being steep descent we can limit the integration by the small part of the contour for the price of an exponentially small correction 
\begin{eqnarray}\label{key}
I_{N,p,2}=\int_{-\epsilon+\mathrm{i}a}^{\epsilon+\mathrm{i}a} e^{N\left( h(z^*e^{\mathrm{i}\varphi})+ (\rho_c-\rho)(\log z^*+\mathrm{i}\varphi)\right)}\frac{z^*e^{\mathrm{i}\varphi}}{(1-e^{\mathrm{i}\varphi})^2}\frac{d\varphi}{2\pi}+O(e^{-\delta N}),
\end{eqnarray}
where for the function $h(z)$ we take the one corresponding to the critical density $h(z)=\log F(z)-\rho_c\log z.$
Next, we approximate the integrand using expansions  at $\varphi=0$
\begin{eqnarray}
\!\!\!\!\!\!\!\!e^{N h(z^*e^{\mathrm{i}\varphi})+ (\rho_c-\rho)(\log z^*+\mathrm{i}\varphi)}\frac{z^*e^{\mathrm{i}\varphi}}{(1-e^{\mathrm{i}\varphi})^2}=\\
-\frac{z^* e^{N \left(h(z^*)+ (\rho_c-\rho)(\log z^*)\right)-\frac{\varphi^2}{2}Nh_2+\mathrm{i}\beta\sqrt{Nh_2}\varphi}} {\varphi^2}\left(1-\frac{\mathrm{i}h_3\varphi^3 N}{6}\right)\nonumber
\end{eqnarray}
where  $h_2=\rho_c(1-\rho_c),$ and $h_3=h_2(1-2\rho_c),$ and the difference $(\rho_c-\rho)$ in the exponent was absorbed into $\beta$ defined in (\ref{eq: beta_def}).  Using  inequality $|1+x-e^{x+y}|<|y|+|x+y|^2e^{|x+y|}$
and choosing $\epsilon$ small enough one can bound the error coming from this approximation by 
\begin{equation}\label{eq: correction to I,p,2}
e^{N \left(h(z^*)+ (\rho_c-\rho)(\log z^*)\right)-C \frac{\varphi^2}{2}Nh_2}\left(O(N^2\varphi^4)+O(N\varphi^2)+O(1))\right)
\end{equation}
for some  $C>0$. Then, going to  rescaled variables $\tilde{\varphi}=(\sqrt{h_2N}\varphi-\mathrm{i}\beta)$, choosing $a=(\alpha+\beta)/\sqrt{Nh_2} $ with some finite $\alpha>-\beta $ and sending $N$ in the limits of integration 
to infinity for the price of another exponentially small correction we arrive at 
\begin{equation}
\!\!\!\!\!\!\!\!\!\!\!\!\!\!\!\!	I_{N,p,2}=-\sqrt{Nh_2}z^* e^{N \left(h(z^*)+ (\rho_c-\rho)\log z^*\right)-\beta^2/2}\left[\mathrm{i}I'(\beta)+\frac{h_3\beta}{6h_2^{3/2}\sqrt{2\pi N}}+O\left(\frac{1}{N}\right)\right],
\end{equation}
where the main part is given in terms of the derivative of function 
\begin{equation}
I(\beta) = \frac{1}{2\pi} \int_{-\infty+\mathrm{i}\alpha}^{+\infty+\mathrm{i}\alpha} \frac{e^{-\frac{ \varphi^2}{2}} }{(\varphi + i\beta)} d \varphi= - \frac{i}{2} \ {\rm erfc}(\frac{\beta}{\sqrt{2}}) e^{\frac{\beta^2}{2}}
\end{equation}
which is in turn related to the function $\mathcal{F}(\beta)$ introduced in (\ref{j_N_sc_function}), 
\begin{equation}
\mathcal{F}(\beta)=-\mathrm{i}\sqrt{2\pi}I'(\beta),
\end{equation}
and the $O(1/N)$ correction have appeared from integration of (\ref{eq: correction to I,p,2}). Using similar
asymptotic expansion of $Z(N,p)$,
\begin{equation}\label{eq: Z_{N,P} crossover}
Z(N,p)=\frac{ e^{N \left(h(z^*)+ (\rho_c-\rho)\log z^*\right)-\beta^2/2}}{\sqrt{2\pi Nh_2}}\left[1-\frac{h_3\beta(\beta^2-3)}{6h_2^{3/2}\sqrt{ N}}+O\left(\frac{1}{N}\right)\right],
\end{equation}
we arrive at the asymptotic formula for  the current 
\begin{eqnarray}\label{eq: j crossover two orders}
 j_N \left(\rho\right)&=&  (R\rho_c+L(1-\rho_c))	N\\
&\times& \left(\mathcal{F}(\beta)+\frac{1-2\rho_c}{6\sqrt{N\rho_c(1-\rho_c)}} 
\beta \left( \mathcal{F}(\beta)(\beta^2-3) -1\right) + O\left(\frac{1}{N}\right)\right),\nonumber
\end{eqnarray}
where in the l.h.s. we still imply that $\rho=\rho_c-\beta\sqrt{\rho_c(1-\rho_c)/N}.$
Thus, the leading order of $O(N)$ confirms  (\ref{eq: j crossover}). We also have obtained the $O(\sqrt{N})$ finite size correction, which is expressed in terms of the same scaling function $\mathcal{F}(\beta)$ and vanishes at the critical point in agreement with (\ref{eq: j_cr}).

Similar analysis is to be done for the diffusion coefficient starting with the formula (\ref{eq: Delta L}).
Here, we focus only on the leading order terms. To identify corresponding terms in the integrand we observe from the above calculation that 
each  factor with the  $z_1$ pole in the denominator contributes the factor $\sqrt{N}$ to the final answer.  Separating in this way   the most singular terms in the integrand of $\Delta^L$ we obtain 
\begin{eqnarray}
\!\!\!\!\!\!\!\!\!\!\!\!\!\!\!\!\!\!\!\!\!\!\!\!\!\!\!\!\!\!\!\!\!\!\!\!\Delta^L&=&N^2\frac{Z(2N,2p)}{Z(N,p)^2}\oint D_{2N,2p}(t)  a^L(t)  \\
\!\!\!\!\!\!\!\!\!\!\!\!\!\!\!\!\!\!\!\!\!\!\!\!\!\!\!\!&+&N^2\oint\oint D_{N,p}(t) D_{N,p}(y) ty \left((1-q)q^2\frac{2+qt+qy}{(1+qy)^2(1+qt)^2} -2 \frac{j_N^L}{\rho(1+qy)} \frac{q}{(t-qy)}\right)+O(N^3),\nonumber
\end{eqnarray}
where   terms we explicitly shown  yield  $O(N^{7/2})$ contribution, while the correction is of order of  $O(N^3)$. Surprisingly, the double integral in the second line cancels in the leading $O(N^{7/2})$ order yielding another  $O(N^3)$ correction. Thus, similarly to sub-critical regime the only contribution to the dominant asymptotics  comes from the integral in the first line, which can be exactly evaluated using (\ref{eq: j_2N-j_N}), while the asymptotics  follow from (\ref{eq: Z_{N,P} crossover},\ref{eq: j crossover two orders}). A  similar calculation  works for $\Delta^R$ yielding  the same result times $q$. This completes the derivation of (\ref{eq: crossover_dif_coef},\ref{eq: Sc_fun_dif_coef}).

\appendix
\section{The particle current as a stationary state observable \label{sec: stationary current}}
It was noted in section \ref{sec: AAP:Model and results} that unlike the higher scaled  cumulants of the particle current in AAP  the first one, the average current, is the stationary state observable, i.e. can be obtained from averaging over the stationary distribution. The observable to average is however highly non-local due to the avalanche dynamics. One of the ways to cope with the non-locality 
is to consider the discrete time model, from which the continuous time AAP would be obtained as a limit. It was shown in \cite{PM2006} that the integrable discrete time zero-range process, known as a q-boson process, looked at from the moving reference frame can be used as such a model.  Here we take another route undertaken in \cite{PPC2003} by considering a discrete time version of the AAP itself, which being non-integrable still has the integrable continuous time AAP as a limit. 

Let us define the discrete time AAP (DAAP) as a particle system evolving on a ring of size $N$.   Unlike the continuous time model,  the state space of  DAAP, $\Omega_{DAAP} \subset \mathbb{Z}^{\mathbb{Z}/N\mathbb{Z}}$, consists of particle configurations $\mathbf{n}=(n_1,\dots,n_N)$, in which  more than one  particle in a site is allowed. In fact, the dynamics allows at most one site $i$ such that $n_i>1$, which will be referred to as active.  
\begin{equation}
\Omega_{DAAP}=\{\mathbf{n}=(n_1,\dots,n_N) \in \mathbb{Z}^{\mathbb{Z}/N\mathbb{Z}}: |\{i\in\{1,\dots,N\}:n_i>1\}|\leq 1\}
\end{equation}
Starting form an initial configuration $\mathbf{n}(0)\in\Omega_{DAAP}$ the subsequent evolution $\mathbf{n}(t)$ is as follows. 

\textbf{If there is no an active site,} $n_k(t) \leq 1,$ for any $k=1,\dots, N$, one of the following options is realized
\begin{itemize}
	\item one of particles, say from site $i$, jumps to an empty neighbouring site: right with probability $\frac{R\delta}{p}$ or left with probability $\frac{L\delta}{p}$;
	\item if the  site $(i+1)$  was occupied before the right jump, $n_{i+1}(t)=1$, then the site becomes active with $n_{i+1}(t+1)=2$ and $n_i(t+1)=0$;
	\item  if the  site $(i-1)$   was occupied before the left jump,  $n_{i-1}(t)=1$, 
	\begin{itemize}
		\item   either with probability $\mu_2$ the two particles from that site immediately move right together, so that the  initial site $i$ becomes active with $n_i(t+1)=2$ and $n_{i-1}(t+1)=0$,
		\item or one of the two particles stays and the other jumps right with probability $1-\mu_2$, so that the initial configuration remains unchanged.   
	\end{itemize}
\end{itemize}

\textbf{If there is an active site $i$ with $n_i(t)\geq 2$ particles for some $i=1,\dots, N$}
\begin{itemize}
	\item either all $n_i(t)$ particles move from site $i$ to site $i+1$ with probability $\mu_{n}$, i.e. $n_i(t+1)=0$ and $n_{i+1}(t+1)=n_{i+1}(t)+n_{i}(t)$,
	\item  or $n_i(t)-1$ particles move from site $i$ to site $i+1$ with probability $1-\mu_n$, i.e.  $n_i(t+1)=1$ and $n_{i+1}(t+1)=n_{i+1}(t)+n_{i}(t)-1$.
\end{itemize}
The other sites do not change. 

These dynamical rules  preserve the total number of particles in the system, which will be fixed to  
\begin{equation}
|\mathbf{n}(t)|=\sum_{i=1}^Nn_i(0)=p.
\end{equation}
To  consider a limit to the continuous time AAP in the end,  we consider the evolution of DAAP that starts from an initial configuration $\mathbf{n}(0)\in \Omega_{DAAP}$ without an active site, i.e.  $n_i(0)\leq 1$, $i=1,\dots L$. This  implies $p\leq N$.

The probability $P_t(\mathbf{n})$ for the system to be in a state $\mathbf{n}$ at time step $t$ solves  Chapmen-Kolmogorov equation
\begin{eqnarray}
P_{t+1}(\mathbf{n}) =  \sum_{\boldsymbol{n' \in \Omega_{DAAP}}}  p(\boldsymbol{n'}\rightarrow \mathbf{n}) P_t(\boldsymbol{n'})
\end{eqnarray}
where $p(\boldsymbol{n'} \rightarrow \mathbf{n})$ is the probability of transition from the state $\boldsymbol{n'}$ to the state $\mathbf{n}$ that stems from  the above dynamical rules,  
\begin{eqnarray}
\!\!\!\!\!\!\!\!\!\!\!\!\!\!\!\!\!\!\!\!\!\!\!\!\!\!\!\!\!\!	p(\boldsymbol{n'}\rightarrow \mathbf{n}) = \left\{ 
\begin{array}{ll}
\frac{R \delta}{p}, \quad & \textrm{if} \quad \exists i: \boldsymbol{n'}- \mathbf{n} = e_i-e_{i+1}, \ n'_{i+1} = 0, \ \forall n_i' \leq 1;\\
\frac{L \delta}{p}, \quad & \textrm{if} \quad \exists i: \boldsymbol{n'}- \mathbf{n} = e_{i+1}-e_{i}, \ n'_{i} = 0, \ \forall n_i' \leq 1;\\
(R+L\mu_2)\frac{ \delta}{p}, \quad & \textrm{if} \quad \exists i: \boldsymbol{n'}- \mathbf{n} = e_{i}-e_{i+1}, \ n'_{i} = n'_{i+1}=1, \ \forall n_i' \leq 1;\\
1-\delta \quad & \textrm{if} \quad \boldsymbol{n'}= \mathbf{n}; \  \forall n'_i \leq 1\\
\mu_{n'_i}; \quad & \textrm{if} \quad \exists ! i: n'_i \geq 2, \ \boldsymbol{n'}- \mathbf{n}=n'_i e_i;\\
1-\mu_{n'_i}; \quad & \textrm{if} \quad \exists ! i: n'_i \geq 2, \ \boldsymbol{n'}- \mathbf{n} = (n'_i-1) e_i;\\
0, \quad & \textrm{otherwise}.
\end{array}
\right.
\end{eqnarray}
where $e_i=(0,\dots,0,1,0,\dots,0)$ is a standard unit vector.
Then the stationary state distribution solves the  balance  equation
\begin{eqnarray}\label{balance_equation}
P_{st}(\mathbf{n}) =  \sum_{\boldsymbol{n'}}  p(\boldsymbol{n'}\rightarrow \mathbf{n}) P_{st}(\boldsymbol{n'}).
\end{eqnarray}
The solution of \eqref{balance_equation} can be found as a product of one-site factors \cite{E2000}
\begin{equation}
P(\mathbf{n}) = \frac{1}{Z_d(N,p)} \prod_{i =1}^L f(n_i),
\end{equation}
where 
\begin{equation}
Z_d(N,p)=\sum_{n\in\Omega_{DAAP}}\prod_{i =1}^L f(n_i)
\end{equation} 
is the normalization constant. The one-site factors can be shown  to  satisfy the recurrent relations 
\begin{eqnarray}
f(n+1) = \frac{\mu_n}{1-\mu_{n+1}}\frac{f(1)}{f(0)} f(n),  \quad n\geq 2,
\end{eqnarray}
and
\begin{eqnarray}
f(2) = \frac{\delta}{p}\frac{(R+L \mu_2)}{1-\mu_2}\frac{f(1)}{f(0)} f(1).
\end{eqnarray}
Fixing the multiplicative constants to $f(0)=f(1)=1$ we obtain  
\begin{eqnarray} \label{eq: f(n)}
f(n+1) = \frac{\delta}{p}\prod_{i=2}^n \frac{\mu_i}{1-\mu_{i+1}} \frac{(R + L \mu_2)}{1-\mu_2}, \quad n \geq 1.  
\end{eqnarray}
The factorized form of the stationary state measure in  DAAP simplifies greatly the calculation of   the averages of observables over the stationary state.  In particular the  mean particle current $J_{DAAP}$ can be obtained as follows. For every site we introduce the random variable  $j_i$ equal to the number of jumps out of the site  $i$ and separate the terms according to the number of particles $n_{i}$ in the  departure sites 
\begin{eqnarray} \label{eq: J_via_P(n)}
J_{DAAP} = \sum_{\mathbf{n},i|n_{i} = 1} P(\mathbf{n}) \mathbb{E}[j_i|n_{i} = 1] + \sum_{\mathbf{n},i| n_{i} \geq 2} P(\mathbf{n}) \mathbb{E}[j_i|n_{i} \geq 2].
\end{eqnarray}
The mean local instantaneous particle current out of a site conditioned to the  site being active or not  is  
\begin{eqnarray} \label{eq:exp_j_1}
\mathbb{E}[j_i|n_{i} = 1] = \frac{\delta \cdot R }{p} - \frac{ \delta \cdot L }{p} \mathbbm{1}_{n_{i-1} = 0} + \frac{\delta \cdot L \mu_2}{p} \mathbbm{1}_{n_{i-1} = 1},\\
\label{eq:exp_j_2}
\mathbb{E}[j_i|n_{i} \geq 2] = n_{i} \mu_n + (n_{i}-1) (1-\mu_n),
\end{eqnarray}
where we use the indicator function to separate the one-step avalanches from all the other avalanches started from the first left jump. These formulas are valid for  general probabilities $\mu_n$. Substituting  the probabilities \eqref{integr_cond}  ensuring the Bethe ansatz  integrability of AAP into  \eqref{eq: f(n)}- \eqref{eq:exp_j_2} and simplifying the result  we obtain 
\begin{eqnarray}\nonumber
J_{DAAP} = \frac{\delta}{pZ_d(N,p)} \Big( Rp C_N^p + L N(\mu_2 C_{N-2}^{p-2} - C_{N-2}^{p-1}) +\\ +(R+L\mu_2) N \sum_{ k=2}^p  C_{N-1}^{p-k} \left(\frac{(-q)^{k-1}k}{k_q} + \frac{(-q)^{k-2}(k-1)}{[k-1]_q}\right)\Big). \label{J_prel}
\end{eqnarray}
This formula can be recast in the integral form. Specifically representing the binomial coefficients as integrals with the use of the generating function (\ref{eq: F(z)}) of stationary weights of continuous time AAP we transform (\ref{J_prel}) to
\begin{eqnarray}
J_{DAAP} = \frac{\delta}{p} \frac{N (1-q)}{Z_d(N,p)} \oint \frac{F^{N}(z)}{z^{p}}  \Big[R g'(zq) - L g'(z) \Big]\frac{dz}{2\pi i}.\label{J_int_form}
\end{eqnarray}
Here we used $g'(z) = g'(qz) - (1-q)^{-1}(1+z)^{-2}$ for the left term. 

One can go from the discrete time to continuous one with the rescaling $t\delta/p\to t$   sending $\delta\to0$. In particular for the current we have 
\begin{equation}
J = \lim_{\delta \to 0} \frac{pJ_{DAAP}}{\delta},
\end{equation} 
which  with the use of 
\begin{equation}
Z_d(N,p) = Z (N,p) +O(\delta)
\end{equation}
yields the result (\ref{eq: j_Nresult_int_form},\ref{eq:j_R,j_L def}).

\end{document}